\pdfoutput=1
\documentclass[12pt]{iopart}

\usepackage[utf8]{inputenc}
\usepackage{lipsum}
\usepackage{graphicx}
\usepackage{pgfplots}

\expandafter\let\csname equation*\endcsname\relax
\expandafter\let\csname endequation*\endcsname\relax
\usepackage{amsmath,amssymb}

\usepackage{listings}
\usepackage{xcolor}
\usepackage{tikz}
\usepackage{bm}
\usepackage{randomwalk}
\usepackage{bm}
\pgfplotsset{width=10cm, height=6cm}
\usetikzlibrary{graphs,graphs.standard}
\usetikzlibrary{decorations.pathreplacing}
\usetikzlibrary{patterns}

\definecolor{dgreen}{rgb}{0,0.7,0}

\def\bluew#1{{\color{blue} #1}}

\usepackage{hyperref}

\newcommand{\Lathrop}[9]{
\begin{scope}
\pgfmathsetmacro{\picwidth}{#1*#2}
\ifthenelse{\equal{#8}{y}}
    {\draw[#9] (0,#6) -- (\picwidth,#6) (0,#7) -- (\picwidth,#7);}
    {}
\draw[#4] (2.55,0.62)
\foreach \x in {1,...,#1}
{   -- ++(#2,rand*#3)
}
coordinate (#5) ;
\end{scope}
\node[right,#4] at (#5) {#5};
}

\usepackage{amsfonts}
\usepackage{graphicx}
\usepackage{hyperref}

\usepackage{xcolor}
\usepackage{subfigure}
\usepackage{textcomp}
\usepackage{cite}
\usepackage{float}
\usepackage{soul}
\usepackage{stackrel}
\usepackage{graphicx}
\graphicspath{{figures_arap/}}

\pdfoutput=1
\usepackage{esint}

\usepackage{color}
\definecolor{dgreen}{rgb}{0,0.7,0}
\def\bluew#1{{\color{red} #1}}

\def\bluew#1{{\color{black} #1}}
\def\blueww#1{{\color{black} #1}}
\newcommand{\beq}{\begin{equation}}
\newcommand{\eeq}{\end{equation}}
\newcommand{\bea}{\begin{eqnarray}}
\newcommand{\eea}{\end{eqnarray}}

\begin{document}

\title[]{Emerging cost-time Pareto front for diffusion with stochastic return}

\author{Prashant Singh}

\address{Niels Bohr International Academy, Niels Bohr Institute,
University of Copenhagen, Blegdamsvej 17, 2100 Copenhagen, Denmark}
\ead{prashant.singh@nbi.ku.dk}

\vspace{10pt}

\begin{abstract}
Resetting, in which a system is regularly returned to a given state after a fixed or random duration, has become a useful strategy to optimize the search performance of a system. While earlier theoretical frameworks focused on instantaneous resetting, wherein the system is directly teleported to a given state, there is a growing interest in physical resetting mechanisms that involve a finite return time. However employing such a mechanism involves cost and the effect of this cost on the search time remains largely unexplored. Yet answering this is important in order to design cost-efficient resetting strategies. Motivated from this, we present a thermodynamic analysis of a diffusing particle whose position is intermittently reset to a specific site by employing
a stochastic return protocol with external confining trap. We show for a family of potentials $U_R(x) \sim |x|^{m}$ with $m>0$, it is possible to find optimal potential shape that minimises the expected first-passage time for a given value of the thermodynamic cost, \emph{i.e.} mean work. By varying this value, we then obtain the Pareto optimal front, and demonstrate a trade-off relation between the first-passage time and the work done.
\end{abstract}

\section{Introduction}
Over the past decade, stochastic resetting has become a prominent area of research in statistical physics. Resetting, in which a system is regularly returned to a given state after a fixed or random duration, is a useful strategy to obtain non-equilibrium properties and optimize search operations \cite{PhysRevLett.106.160601, Evans_2011_1, PhysRevE.91.012113, PhysRevLett.118.030603}. In the past few years, the subject has been rigorously studied across cross-disciplinary disciplines including physics \cite{PhysRevE.91.052131, PhysRevE.91.012113, PhysRevLett.118.030603, PhysRevLett.113.220602, Bressloff_2020gfs, PhysRevLett.116.170601, Gupta_2019cjd, Singh_20202d, Evans_2018vaf, PhysRevE.102.052129, PhysRevE.103.052119,Stojkoski_2022_auto, Majumdar_2018_auto,Singh_2022aphnx,PhysRevResearch.5.013122}, chemical and biological processes \cite{doi:10.1073/pnas.1318122111, PhysRevE.93.062411, PaulBressloff_2020, PhysRevResearch.3.L032034}, income models \cite{income-Pal, Santra_eco} and computer science \cite{LUBY1993173, HamlinThrasherKeyrouzMascagni+2019+329+340}. Earlier theoretical frameworks mostly focused on instantaneous resetting, wherein the system is directly teleported to a resetting state. \blueww{A commonly studied model is that of a one dimensional diffusing particle whose position $x(t)$ resets to a value $x_R$ at a rate $r$.} In a small time interval $[t, t+\Delta t]$, the position evolves as
\begin{align}
x(t+\Delta t) & = ~x(t) + \Delta t ~\eta(t), ~~\text{with probability } (1-r \Delta t) , \\
& = ~x_R, ~~~~~~~~~~~~~~~~~\text{with probability } r \Delta t ,
\end{align}
where $\eta(t)$ is the Gaussian white noise with zero mean and correlation $\langle \eta(t) \eta(t')  \rangle = 2 D \delta(t-t')$. Here $D$ is the diffusion coefficient which will be set to $1/2$ throughout this paper. During resetting, the position instantaneously changes its value to $x_R$. Several theoretical results have been already established for this model as well as for other models in the past one decade \cite{reset-review1}. For example, this simple strategy gives a finite average first-passage time in models which otherwise have infinite average first-passage time. We refer to \cite{reset-review1, reset-review2, reset-review3, reset-review4, Pal_2022jafda12} and references therein for a thorough review on the topic. While these studies provide useful insights in understanding the ramifications of resetting, they fail to capture the intricacies of real models where resetting is usually non-instantaneous. Single-particle experiments on resetting often involve non-zero return times \cite{reset-Exp1, reset-Exp2, reset-Exp3}. This has naturally prompted some recent attempts to account for the time penalty occuring
during non-instantaneous resetting events \cite{PhysRevResearch.2.043174, PalHT_2019, PhysRevE.100.040101, PhysRevE.101.052130, PhysRevE.100.042104, PaulBressloff_2020, PhysRevE.106.044127,RP-1, RP-2, RP-3, NR-1, NR-2, intermittent-1, intermittent-2}. Some works, for instance, incorporate a residence time following each resetting event \cite{RP-1, RP-2, RP-3}, while others involve intermittently switching on and off certain confining potentials \cite{intermittent-1, intermittent-2}.

Stochastic return dynamics is a useful extension of the latter strategy \cite{trap-1, trap-2, trap-3, trap-4}. In this case, a particle evolves under its natural dynamics. However, one intermittently switches on a confining trap that takes the particle to a desired location. Upon reaching this location, the trap is switched off, and the particle resumes its natural dynamics. Such a protocol also gives rise to a finite and optimised value of the mean first-passage time, in contrast to its diverging value for the simple diffusion \cite{trap-1, trap-2}. However, it is more realistic since the particle takes a non-zero time to reach the resetting location.

Employing the confining potential incurs an energetic cost \cite{STR-1}. According to the principles of stochastic thermodynamics, any temporal change of the external trap is related to the work done on the system \cite{Seifert-review,Sekimoto1998}. Hence, it is quite natural to anticipate that thermodynamics will play a crucial role in resetting problems. Several works have also focused on the thermodynamic aspects of resetting \cite{STR-2,STR-3,STR-4,STR-5,STR-6,STR-7,STR-8,STR-9,STR-10,STR-11,STR-12, STR-13,STR-14, Genthon_2022, STR-rev-1}. \blueww{Recently, the first and second laws of thermodynamics were tested experimentally for a resetting system and the energetic cost was shown to satisfy the  Landauer’s bound \cite{expttt-1}.} It is also important to understand the impact of thermodynamic costs on the search properties. \blueww{For instance, in many biochemical systems that discriminate between different molecules, the reaction can reset whenever a wrong molecule is selected \cite{discr}. However, these reaction pathways often involve consumption of energetic molecules and it is crucial to understand the effect of energetic cost on the turnover time of the right product molecule. Similarly, in protein-folding dynamics, a protein molecule transitions through different folding states \cite{protein-fold2,protein-fold1}. However, it can occasionally get trapped in a misfolded state, delaying the completion time to reach the final state. Molecular chaperones can restore the protein back to its initial unfolded state. But the action of chaperone is driven by ATP hydrolysis. Once again, the completion time of a resetting process strongly depends on the energetic cost driving the process.}

While statistical properties of cost functions and first-passage time for resetting systems are individually well studied, their inter-dependence remains not well explored. Only recently in \cite{NTR-2}, the role of cost functions on the search efficiency was studied for diffusion with  instantaneous resetting. Another recent work \cite{STR-13} considered the interplay between first-passage time and work done for stochastic return model with linear potential. Beyond linear case, this interplay however remains still unexplored. A natural question is - what resetting potential optimises the search time for a given value of the cost function? Answering this might be relevant in designing cost-efficient resetting protocols. 

However, optimising both work and first-passage time is mutually exclusive. A stronger potential rapidly resets the particle, thereby reducing the first-passage time to find the target. However, the cost associated with employing such a stronger potential is quite significant. In contrast, a weaker potential involves lower cost, but the time spent during the return phase is generally higher. In such a scenario of mutually exclusive objective functions, a commonly used optimisation approach is that of the Pareto front \cite{Pareto}. Essentially, one minimises an objective function (say $\mathcal{O}_1$) while keeping the other one (say $\mathcal{O}_2$) fixed. The collection of all optimal solutions corresponding to different values of $\mathcal{O}_2$ then generate the Pareto front. It shows that beyond a certain point it is not possible to optimise $\mathcal{O}_1$ without changing $\mathcal{O}_2$ further, and there is a limit on how much one can optimise $\mathcal{O}_1$.

 \blueww{In this paper, we are interested in carrying out such an optimisation scheme for averages of first-passage time and work of a one dimensional Brownian motion. Focusing on a family of resetting potentials $U_R(x) \sim |x|^{m}$ with $m>0$, we will obtain the Pareto front for these averages and use it to demonstrate a trade-off relation between them. These potentials can be experimentally realized using optical traps \cite{reset-Exp1, reset-Exp2,expttt-1}. Essentially a colloidal particle diffuses under the action of a weak potential. With some rate, a strongly confining potential is switched on that quenches the particle to a desired location. Once the particle has reached this location, the strongly confining potential is switched off and the colloid again diffuses. Such an experimental set-up is well-suited to verify theoretical results derived in this paper. 
Diffusion with resetting has also proven useful in modelling biological systems, such as in RNA polymerase backtracking \cite{PhysRevE.93.062411} or the target search of certain transcription factors \cite{hauba}. Our study might serve as a benchmark for exploring similar questions in these systems. Apart from this, our paper presents a methodology that can be extended to general Markov processes in higher dimensions. The trade-off relation shown in this paper, and in other work \cite{STR-13}, might be useful in probing similar trade-offs in other stochastic processes.}

The remainder of paper is structured as follows: in section \eqref{sec-model}, we introduce our model and fix all notations relevant for the subsequent study. Section \eqref{sec-renewal} presents the derivation of the renewal formula for the joint distribution, which becomes instrumental in deriving the mean work in section \eqref{sec-work} and the mean first-passage time in section \eqref{sec-FPT} for arbitrary resetting potentials. Section \eqref{sec-tradeoff} discusses Pareto optimisation and the trade-off relation, followed by the conclusion in section \eqref{sec-conclusion}.

\begin{figure}[t]
\includegraphics[scale=0.33]{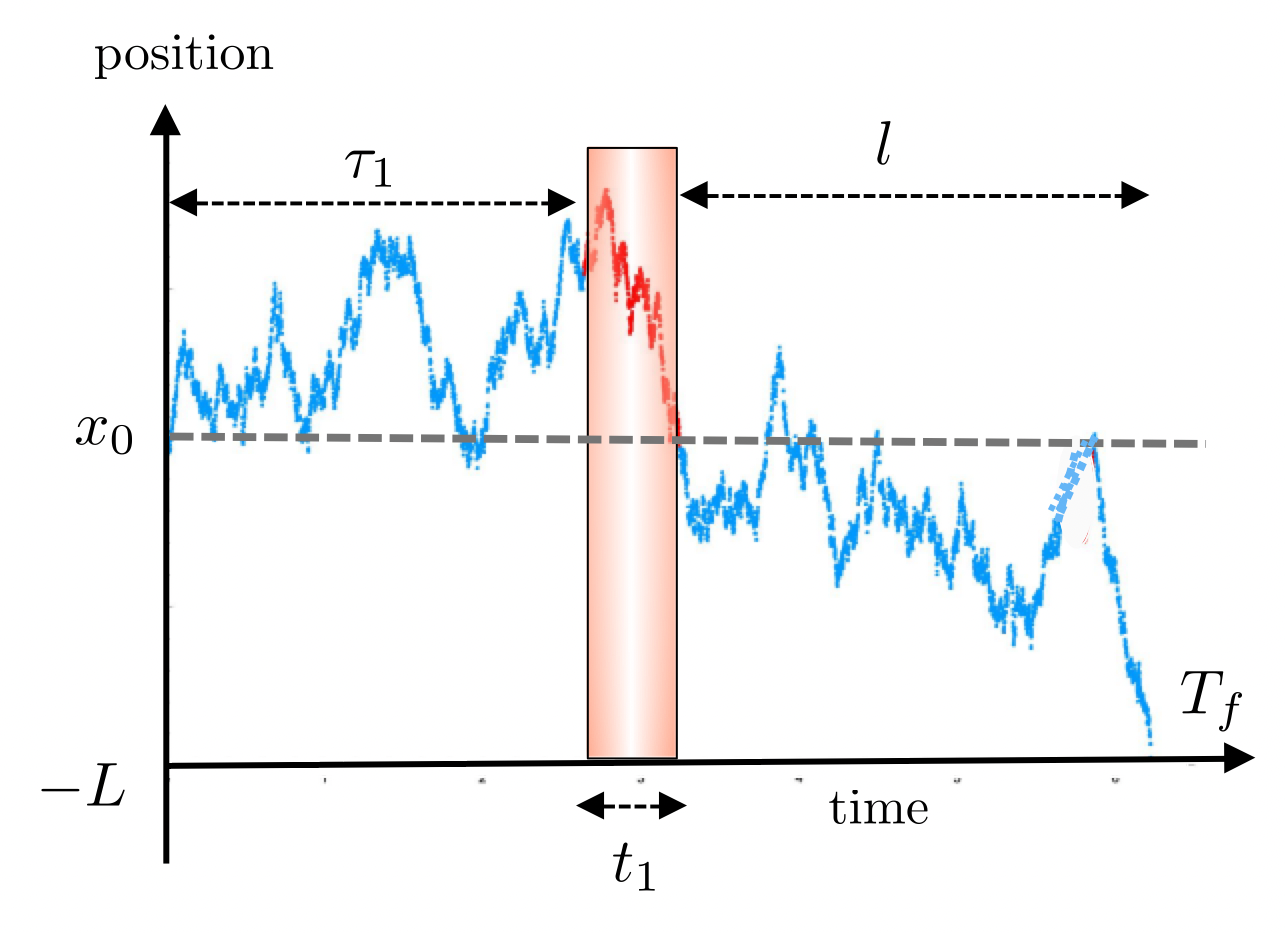}
\includegraphics[scale=0.33]{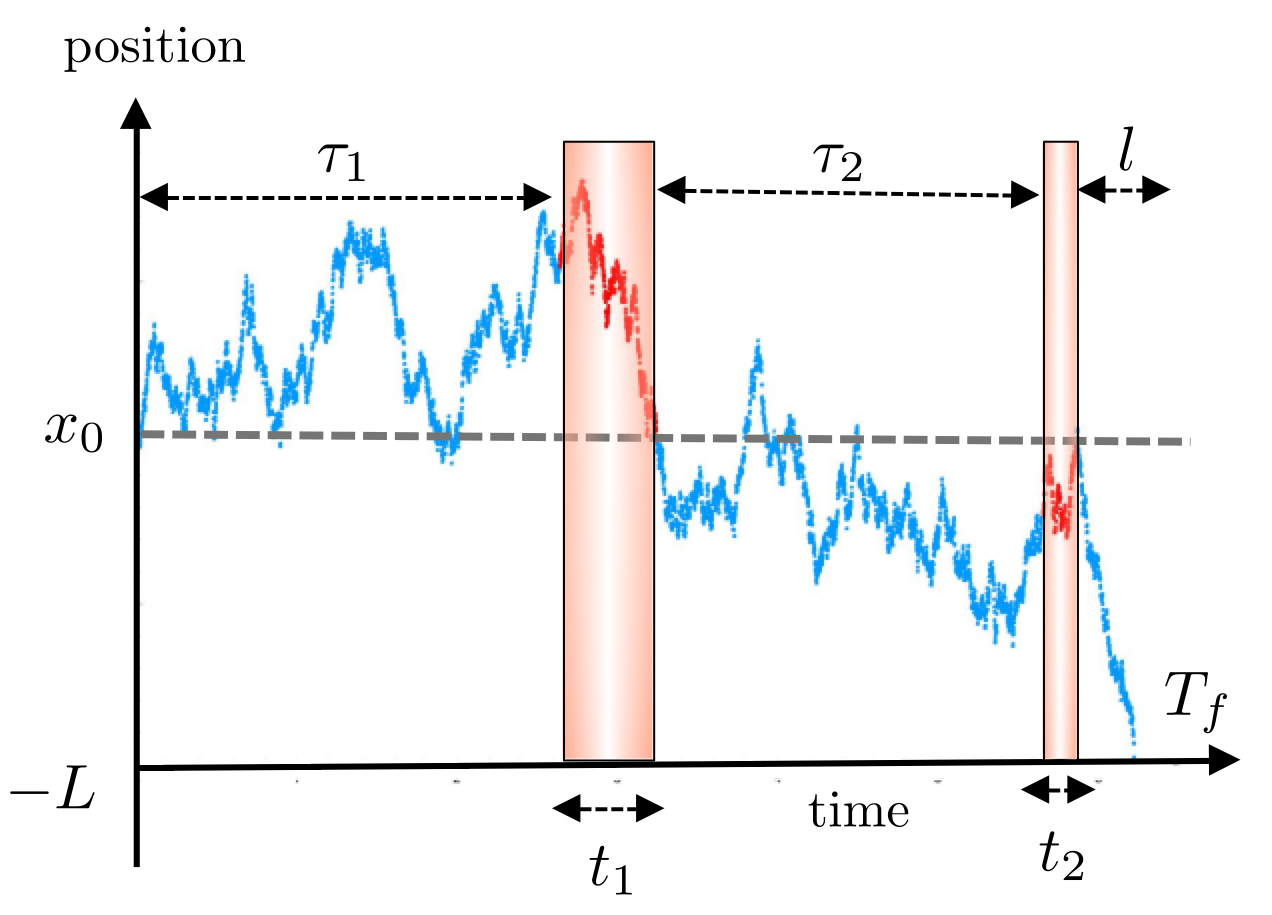}
\centering
\caption{Schematic illustration of a trajectory of a stochastic process undergoing one (left panel) and two (right panel) resetting events at a rate $r$. The resetting events, shown in red within the shaded regions, are implemented using a trapping potential $U_R(x)$ with its minimum located at the origin $x_0 = 0$. During these events, the particle makes first-passage visits to the origin, with the associated first-passage times denoted by $t_1, t_2, \ldots, t_n$ (with $n = 1$ for the left panel and $n=2$ for the right panel). Apart from this, the particle follows its original dynamics (shown in blue) for durations $\tau_1, \tau_2, \ldots, \tau_n$, each drawn from the exponential distribution $\rho(\tau) = re^{-r \tau}$. \blueww{During the last duration $l$, the process reaches the target location at $x=-L$.} The global first-passage time to the position $x= - L$ (with $L>0$) is denoted by $T_f$.
}    
\label{fig-trajectory}
\end{figure}
\section{Model}
\label{sec-model}
Let us consider a freely diffusing particle in one dimension whose position at time $t$ is represented by a continuous variable $x(t)$. \blueww{The process is governed by the equation for overdamped motion}
\begin{align}
\frac{dx}{dt} =  \eta(t), \label{model-eq-1}
\end{align} 
where $\eta(t)$ is the Gaussian white noise as introduced before.  We fix the initial position to the origin, $x_0 = 0$. The particle evolving with this dynamics also experiences resetting at a constant rate $r$. During the resetting event, an external confining potential $U_R(x)$ is switched on such that the particle is facilitated to move towards the minimum of this potential. We assume that this minimum is unique and located at the origin. During the return phase, the position evolves according to
\begin{align}
\frac{dx}{dt} =  -\frac{\partial U_R(x)}{\partial x} + \eta(t). \label{model-eq-21}
\end{align}
This phase, which we will refer to as return/resetting phase, ends whenever the particle reaches the origin for the first time, at which point the potential is switched off. After this, it again evolves according to Eq.~\eqref{model-eq-1} until the next resetting event. Thus, the process evolves in a series of diffusive and resetting phases, and it can be succinctly written as
\begin{align}
\frac{dx}{dt} =  -\frac{\partial V \left( x, \Lambda(t) \right)}{\partial x} + \eta(t),~~~\text{with }V \left( x, \Lambda(t) \right) \equiv \Lambda(t) U_R(x). \label{model-eq-2}
\end{align}
Here $\Lambda(t)$ is a random dichotomous variable which takes value $1$ if the particle is in the return phase and $0$ if it is in the diffusive phase. Notice that for any finite $U_R(x)$, the return phase is non-instantaneous and the particle spends a non-zero time to reach the resetting site. The process \eqref{model-eq-2} is repeated until the particle reaches the target position located at $x(t)= -L$ (with $L>0$) for the first time in the diffusive phase. We denote this global first-passage time by $T_f$, see Figure \ref{fig-trajectory}.

While implementing resetting using trapping potentials, one always incurs some thermodynamic cost, i.e. work \cite{STR-1,STR-13}. In the last few decades, the framework of stochastic thermodynamics has been rigorously developed and this has allowed us to identify  thermodynamic quantities for stochastic systems such as in Eq.~\eqref{model-eq-2} \cite{Seifert-review,Sekimoto1998}. The total stochastic work until the first-passage time $T_f(L)$ is given by
\begin{align}
W(L) = \int _0 ^{T_f} dt ~\dot{\Lambda}(t)  ~\frac{\partial V \left( x, \Lambda(t) \right)}{\partial  \Lambda(t) } = \int _0 ^{T_f} dt ~\dot{\Lambda}(t)  ~U_R(x(t)). \label{pp-eq-1}
\end{align}
\bluew{Physically, this quantity represents the amount of work required to implement the resetting potential until its first-passage event. Hence, it acts as a suitable cost function that gives a measure of energetics involved in the process.} We are interested in calculating the statistical properties of this cost for a general potential $U_R(x)$. To achieve this, we will construct a renewal formula for the joint distribution of the work done and the global first-passage time in terms of the characteristic properties of the underlying processes in Eqs.~\eqref{model-eq-1} and \eqref{model-eq-21}.
Using this formula, we then derive general expressions for the averages of the work done and the first-passage time for any arbitrary confining potential. \bluew{Some of these results were also previously obtained in \cite{PhysRevResearch.2.043174, PhysRevE.101.052130,STR-13}. We provide here an independent derivation. Also going beyond their formal expressions, we  obtain simple and elegant expressions for a family of potentials $U_R(x) \sim |x|^{m}$ with $m>0$. This will, for the first time, allow us to quantitatively characterize how varying potential shape changes these averages. Using these expressions, we will demonstrate a trade-off relation and also calculate their Pareto front and optimal resetting potentials.}

It is important to note that our first-passage event $T_f$ is achieved only during the diffusive phase, and not during the return phase  \cite{PhysRevResearch.2.043174, STR-13}. While this is valid for strongly confining potentials, it will break down if the potential is only weakly confining and $L$ is not large. Here we still adopt this crucial assumption since our primary goal is to construct the methodology to study the Pareto optimisation. This simplification allows us to derive  precise and elegant expressions, which can be used to obtain optimal front and optimal potential shape. It also enables us to provide a thermodynamic interpretation of some non-thermodynamic cost functions studied in the resetting literature. Modifying our framework to relax this assumption is possible, but it makes the problem analytically less tractable. We aim to address this in a future work.

\section{Renewal formula for the joint distribution of $W(L)$ and $T_f(L)$}
\label{sec-renewal}
We first look at the joint distribution $\mathcal{P}(W, T_f)$ of the work $W(L)$ and the first-passage time $T_f(L)$ in presence of resetting. 
\subsection{Number of resetting $n=1$}
To begin with, consider the situation where particle experiences only one resetting event $(n=1)$ before it gets absorbed at $x= -L$. A schematic illustration of such a trajectory is shown in Figure \ref{fig-trajectory} (left panel). Let us break this trajectory in three parts: (i) pre-reset part $[0, \tau _1]$, (ii) reset part $[\tau _1, \tau _1 + t_1]$ and (iii) post-reset part $[\tau _1 + t_1, \tau _1 + t_1 + l]$. In the pre-reset part, the particle, starting from the origin, reaches some position $x(\tau _1)=x_1$ at time $\tau_1$. Moreover, it always stays above the target at $-L$. Therefore, the statistical weight of this part of the trajectory is just the propagator $P_0(x_1, \tau _1;-L)$ of a freely diffusing particle with an absorbing wall located at $x = -L$.

We next examine the reset part. In this phase, the particle begins its motion from $x_1$ and makes a first-passage visit to the origin in duration $t_1$. Therefore, the contribution from this segment is the distribution $F_R(t_1 \,|\, x_1;0)$ of the first-passage time $t_1$ to the origin, given that the particle was at $x_1$ at the start of this segment. The subscript `$R$' in $F_R(t_1 \,|\, x_1;0)$ is used to indicate that this first-passage time is measured for model \eqref{model-eq-21} with resetting potential $U_R(x)$. Finally, we investigate the post-resetting part in Figure \ref{fig-trajectory} (left panel). As clear from this figure, the particle, beginning from the origin, makes a first-passage visit at time $l$ to the target location at $x = -L$. Then, the contribution from this part is also the first-passage distribution $F_0(l\big|0;-L)$ for the diffusion (with no resetting potential).

Since the process is a Markov process for all three segments, the total path weight of the entire trajectory is equal to the product of these three contributions. Hence, we have
\begin{align}
\text{Path weight } \Big| _{n=1}= P_0(x_1, \tau _1;-L) ~\rho(\tau_1)~F_R(t_1 \,|\, x_1;0)~F_0(l\big|0;-L) ~e^{-r l}. \label{pp-eq-2}
\end{align}
We have weighted $P_0(x_1, \tau _1;-L)$ by the resetting time distribution $\rho(\tau_1) = r e^{-r \tau _1}$. Moreover, in the post-resetting phase, the particle does not experience any resetting event till time $l$, the probability of which is $e^{-r l}$. This results in $e^{-r l}$ factor accompanying the last term.

Once the path weight is calculated, we need to supplement it with the correct values of the work and the first-passage time. Calculating $T_f$ is straightforward since it is given by the sum of all times appearing above, i.e., $T_f = \tau _1 + t_1 + l$. On the other, for work, we need to specify $\Lambda(t)$ in Eq.~\eqref{pp-eq-1}. Recall that $\Lambda(t)$ is a dichotomous variable which takes value $1$ if the particle is in the return phase and $0$ if it is in the diffusive phase. Looking at Figure \ref{fig-trajectory}, we can write it as \cite{STR-1}
\begin{align}
\Lambda(t) =  \Theta (t-\tau _1)- \Theta (t-\tau _1-t_1),
\end{align}
where $ \Theta (t-\tau _1)$ stands for the Heaviside theta function. It takes value $1$ if $t > \tau _1$ and $0$ otherwise. Taking the derivative of $\Lambda(t)$ and plugging it in Eq.~\eqref{pp-eq-1}, we obtain $W = U_R(x(\tau _1)) -  U_R(x(\tau _1+t_1))$. Noting that, at the end of any resetting phase, the particle is at the origin, we get $x(\tau _1+t_1) = 0$. Furthermore, the trapping potential $U_R(x)$ has its minimum value located at the origin. Without any loss of generality, we take this minimum value to be zero, $U_R(x(\tau _1+t_1)) = 0$. The stochastic work then becomes
\begin{align}
W = U_R(x_1). \label{pp-eq-3}
\end{align}
Combining this with Eq.~\eqref{pp-eq-2}, we finally obtain the contribution of $n=1$ events to the joint distribution $\mathcal{P}(W, T_f)$ as
\begin{align}
\mathcal{C}_1 & = \int_{-L}^{\infty} dx_1 \int _{0}^{\infty} d \tau _1 \int _{0}^{\infty} d t _1 \int _{0}^{\infty} d l ~ P_0(x_1, \tau _1;-L) ~\rho(\tau_1)~F_R(t_1 \,|\, x_1;0)~F_0(l\big|0;-L) ~e^{-r l} \nonumber \\
& ~~~~~~~~~~~~~~~~~~~~~~~~~~~~~~~~~~~~~~~~~\times \delta \Big( T_f - \tau _1 - t_1 - l\Big) ~\delta \Big( W - U_R(x_1)\Big). \label{pp-eq-4}
\end{align}
The delta functions enforce the correct values of work and first-passage time.
\subsection{Number of resetting $n=2$}
We next calculate the contribution of those events with $n=2$. A schematic illustration of this is shown in Figure \ref{fig-trajectory} (right panel). For this case, we break the trajectory into five parts: (i) pre-first reset part $[0, \tau _1]$, (ii) first reset part $[\tau _1, \tau _1 + t_1]$, (iii) pre-second reset part $[\tau _1 + t_1, \tau _1 +t_1+\tau _2]$, (iv) second reset part $[\tau_1 + t_1 + \tau _2, \tau _1+t_1+\tau _2 +t_2]$, and (v) post-second reset part $[\tau _1 + t_1 +\tau _2 +t_2,\tau _1 + t_1 +\tau _2 +t_2+l]$. For each of these segments, we can calculate the statistical weight using the same physical rationale as before. For example, in segment (i), the particle, starting from the origin, reaches some position $x(\tau _1) = x_1$ at time $\tau _1$ in presence of an absorbing wall at $x= - L$. Contribution of this segment is $P_0(x_1, \tau _1;-L) \rho(\tau _1)$. Similarly, in segment (iii), the process remains above $x= - L$ and reaches the position $x(\tau _1+t_1+\tau _2) = x_2$ given that it started from the origin at the beginning of this segment. This again yields the contribution $P_0(x_2, \tau _2;-L) \rho(\tau _2)$. On the other hand, in segments (ii) and (iv), the particle makes first-passage visits to the origin at durations $t_1$ and $t_2$ starting from $x_1$ and $x_2$ respectively. We have $F_R(t_1 \,|\, x_1;0)$ and $F_R(t_2 \,|\, x_2;0)$ from these parts. Finally, in segment (v), the particle, during the free diffusive phase, reaches the target site $x= - L$ in a time duration $l$ without experiencing any resetting. The weight of this segment is $F_0(l\big|0;-L)e^{- rl}$.

\noindent
Using Markovianity of the process, the total path probability then turns out to be
\begin{align}
\text{Path weight } \Big| _{n=2}= &~ P_0(x_1, \tau _1;-L) ~\rho(\tau_1)~F_R(t_1 \,|\, x_1;0)~P_0(x_2, \tau _2;-L) ~\rho(\tau_2) \nonumber \\
& ~~~~~~~~F_R(t_2 \,|\, x_2;0)~F_0(l\big|0;-L) ~e^{-r l}. \label{pp-eq-5}
\end{align}
Furthermore, we have
\begin{align}
\Lambda(t) =  \Theta (t-\tau _1)- \Theta (t-\tau _1-t_1)+\Theta (t-\tau _1-t_1-\tau_2)-\Theta (t-\tau _1-t_1-\tau _2-t_2),
\end{align}
using which in Eq.~\eqref{pp-eq-1} gives us the value of work
\begin{align}
W = U_R(x_1)+U_R(x_2). \label{pp-eq-6}
\end{align}
By incorporating this in Eq.~\eqref{pp-eq-5}, we derive the contribution of $n=2$ events to the joint distribution $\mathcal{P}(W, T_f)$ as
\begin{align}
\mathcal{C}_2 & = \int_{-L}^{\infty} dx_1 \int_{-L}^{\infty} dx_2 \int _{0}^{\infty} d \tau _1 \int _{0}^{\infty} d t _1  \int _{0}^{\infty} d \tau _2 \int _{0}^{\infty} d t _2 \int _{0}^{\infty} d l ~ P_0(x_1, \tau _1;-L) ~\rho(\tau_1) \nonumber \\
&~~~~~~~~~~~~~~~~~ \times
F_R(t_1 \,|\, x_1;0)~ P_0(x_2, \tau _2;-L) ~\rho(\tau_2)~F_R(t_2 \,|\, x_2;0)~F_0(l\big|0;-L) ~e^{-r l} \nonumber \\
& ~~~~~~~~~~~~~~~~~~~~~~~~\times \delta \Big( T_f - \tau _1 - t_1-\tau _2 - t_2 - l\Big) ~\delta \Big( W - U_R(x_1)-U_R(x_2)\Big). \label{pp-eq-4}
\end{align}
Proceeding similarly, one can write the contributions $\mathcal{C}_3,~\mathcal{C}_4, \ldots $ to the joint distribution due to $n=3,4,\ldots $ number of resetting events. In particular, for the general value of $n \geq 1$, one gets
\begin{align}
\mathcal{C}_n = & \int _{0}^{\infty} dl ~\left[  \prod _{i=1}^{n}    \int _{-L}^{\infty} dx_i \int _{0}^{\infty} d\tau _i \int _{0}^{\infty} dt _i  ~P_0(x_i, \tau _i;-L) ~\rho(\tau_i)~ F_R(t_i \,|\, x_i;0) \right] \nonumber \\
& \times F_0(l\big|0;-L) e^{-r l}~ \delta \Big(T_f-l-\sum_{j=1}^{n} \left( \tau _j + t_j\right) \Big)~\delta \Big(W-\sum _{j=1}^{n} U_R(x_i) \Big), \label{pp-eq-44}
\end{align}
whereas for $n=0$, where no resetting occurs, one has
\begin{align}
\mathcal{C}_0 =  F_0(T_f \big|0;-L)~ e^{-r T_f}~\delta (W). \label{pp-eq-445}
\end{align}
Since the external potential is not switched on, the value of work for this case is equal to zero.
\subsection{Joint distribution $\mathcal{P}(W, T_f)$}
We now have all contributions essential for computing $\mathcal{P}(W, T_f)$. Summing all of them
\blueww{\begin{align}
\mathcal{P}(W, T_f) =  \sum _{n=0}^{\infty} \mathcal{C}_n. \label{pp-eq-7}
\end{align}
Each of these contributions involves delta function over both $T_f$ and $W$. To get rid of them, it is useful to take Laplace transformations with respect to $T_f~(\to s)$ and $W~(\to k)$. Given the convolution structure of $\mathcal{C}_n$ in Eq.~\eqref{pp-eq-44}, its Laplace transform takes a simple form
\begin{align}
\int _{0}^{\infty}dT_f~e^{-sT_f} \int _0^{\infty}dW~e^{-kW} ~\mathcal{C}_n = & \left[  r \int _{-L}^{\infty} dx ~e^{-k U_R(x)}  \bar{P}_0(x, s+r;-L)~\bar{F}_R(s \,|\, x;0)   \right]^n \nonumber \\
& ~~~~~~\times \bar{F}_0(s+r \,|\, 0;-L),  \nonumber
\end{align}
where $\bar{F}_0(s \,|\, 0;-L)$, $\bar{F}_R(s \,|\, x;0)$ and $\bar{P}_0(x, s;-L)$ are respectively the Laplace transforms of $F_0(t \,|\, 0;-L)$, $F_R(t \,|\, x;0)$ and $P_0(x, t;-L)$. This transformation holds true for all positive integer values of $n$ including zero. Utilizing it in Eq.~\eqref{pp-eq-7} yields
\begin{align}
\mathcal{Z}(k,s) =  \bar{F}_0(s+r \,|\, 0;-L)~\sum _{n=0}^{\infty} \left[  r \int _{-L}^{\infty} dx ~e^{-k U_R(x)}  \bar{P}_0(x, s+r;-L)~\bar{F}_R(s \,|\, x;0)   \right]^n, \nonumber
\end{align}
where $\mathcal{Z}(k,s)$ denotes the double Laplace transformation of $\mathcal{P}(W, T_f)$. If the term inside bracket $[...]$ is less than one, the summation above converges, and we can express it as
\begin{align}
\mathcal{Z}(k,s) =  \frac{\bar{F}_0(s+r \,|\, 0;-L)}{1-r \int _{-L}^{\infty} dx ~e^{-k U_R(x)}~\bar{P}_0(x, s+r;-L)~\bar{F}_R(s \,|\, x;0)  }, \label{pp-eq-9}
\end{align}
This renewal formula is the first main result of our paper.} It gives the exact joint distribution in terms of the underlying propagators and the first-passage properties. 

Our formula also gives the correct normalisation. To see this, we put $k=0$ and $s=0$ in Eq.~\eqref{pp-eq-9}. We also use the normalisation property of the distribution $\bar{F}_R(s=0 \,|\, x;0) = 1$. One then has
\begin{align}
\mathcal{Z}(k=0,s=0) =  \frac{\bar{F}_0(r \,|\, 0;-L)}{1-r \int _{-L}^{\infty} dx ~\bar{P}_0(x, r;-L)}. \label{pp-eq-10}
\end{align}
We next identify that the integration $\int _{-L}^{\infty} dx ~P_0(x, t;-L)$ simply represents the probability $Q_0(t;-L)$ that the particle, starting from the origin, has survived the absorbing wall at $x=-L$ till time $t$. The corresponding first-passage distribution is expressed as the first derivative of the survival probability $F_0(t \,|\, 0;-L) = - d Q_0(t;-L) / dt $ \cite{redner}. In terms of the Laplace variable, these relations translate to
\begin{align}
\int _{-L}^{\infty} dx ~\bar{P}_0(x, s;-L) = \bar{Q}_0(s;-L),~\text{and }\bar{F}_0(s \,|\, 0;-L) = 1-s\bar{Q}_0(s;-L). \label{pp-eq-11}
\end{align}
Substituting them in Eq.~\eqref{pp-eq-10}, we obtain the correct normalisation $\mathcal{Z}(k=0,s=0)  = 1$. This further validates our renewal formula. In what follows, we will use it to study the statistics of $W$ and $T_f$ for a general resetting potential.

\begin{figure}[t]
\includegraphics[scale=0.37]{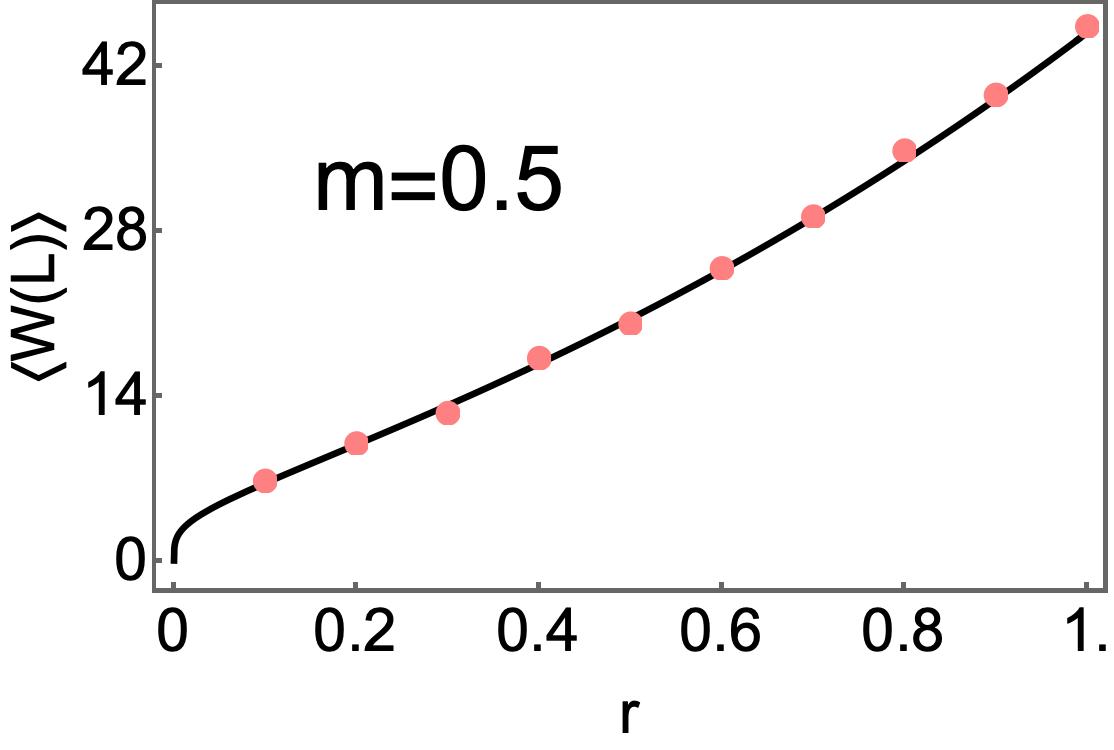}
\includegraphics[scale=0.38]{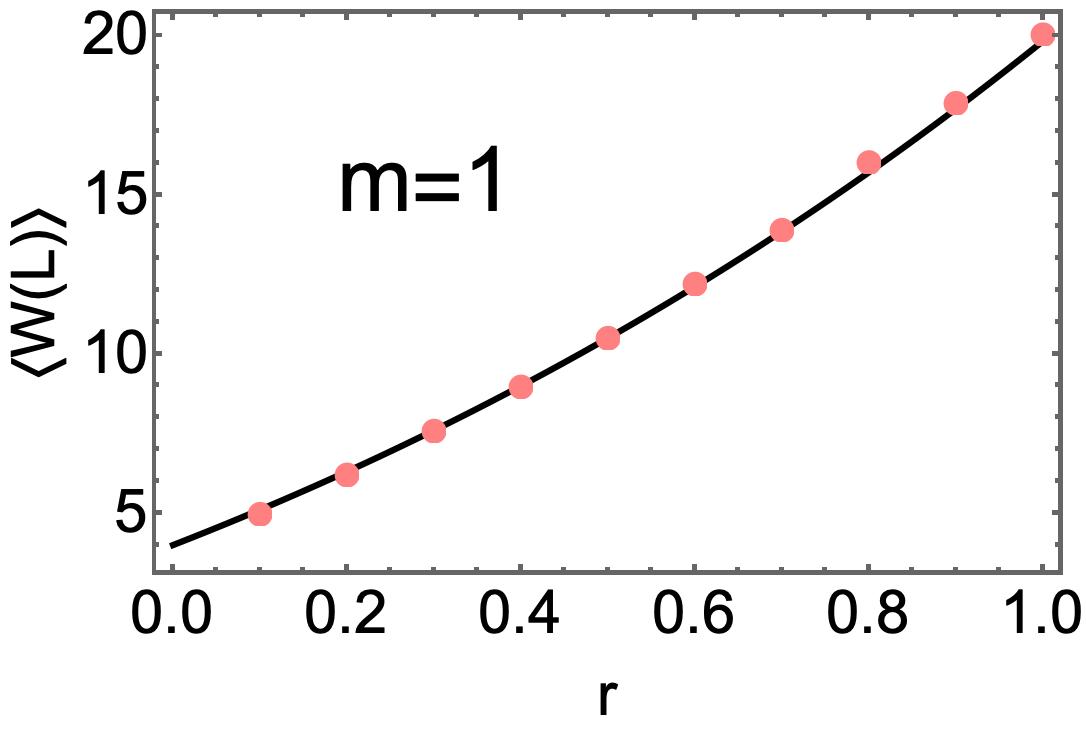}
\includegraphics[scale=0.42]{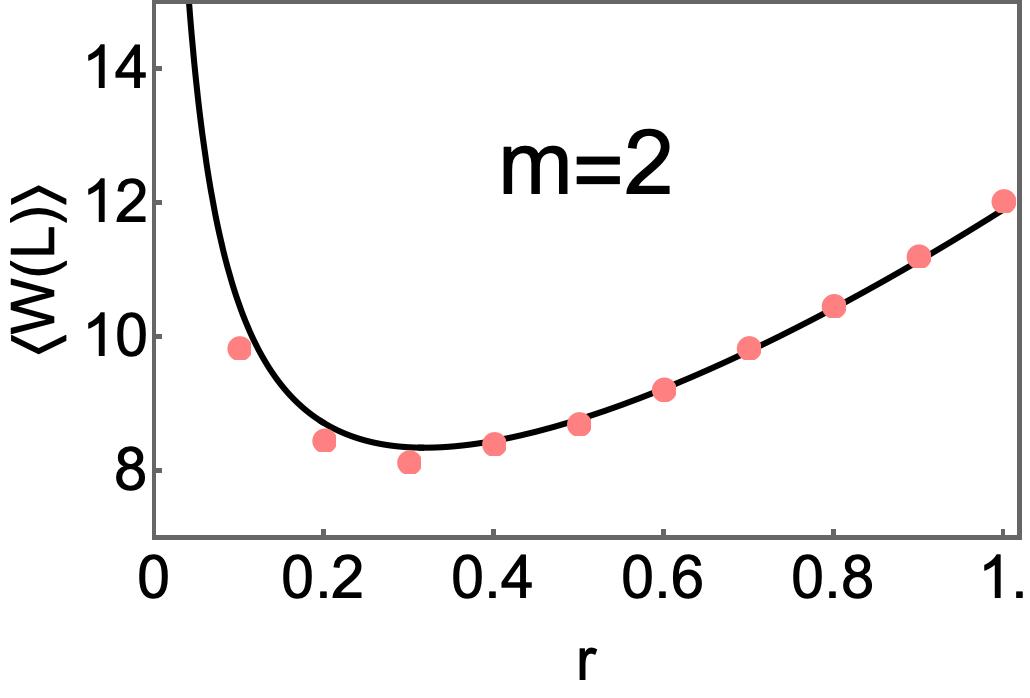}
\centering
\caption{Numerical verification of the mean thermodynamic work for a class of resetting potentials $U_R(x) = \alpha  |x|^{m} /m $. For all three panels, solid lines represent the analytic formula in Eq.~\eqref{pp-eq-17} while symbols are the simulation data. Parameters used for the comparison are $L = 2,~\alpha =2$.
}    
\label{fig-mean-work}
\end{figure}
\section{Work for a general resetting potential}
\label{sec-work}
We will first use Eq.~\eqref{pp-eq-9} to analyze the statistics of work. Setting \(s=0\) and applying the normalization \(\bar{F}_R(s=0 \,|\, x;0) = 1\) gives us the generating function of \(W(L)\) as
\begin{align}
\mathcal{Z}(k,s=0) =  \frac{\bar{F}_0(r \,|\, 0;-L)}{1-r \int _{-L}^{\infty} dx ~e^{-k U_R(x)}~\bar{P}_0(x, r;-L)  }~. \label{pp-eq-12}
\end{align}
Interestingly, our expression matches with the renewal equation for some other non-thermodynamic cost functions studied in the context of instantaneous resetting processes \cite{NTR-1,NTR-2}. In such a set-up, the particle undergoes instantaneous resetting from some position $x'$ to the origin. If one defines $U_R(x')$ as the cost for each resetting event, then the renewal formula for this cost till its first-passage time is exactly the same as Eq.~\eqref{pp-eq-12}. If we integrate out the return time in our problem and focus only on the work done, it becomes mathematically equivalent to \cite{NTR-1,NTR-2}. However, the first-passage time properties will be different, as we show below. We especially emphasize that these works deal with idealized instantaneous resetting, while our derivation is for non-instantaneous resetting implemented through external confining traps.


Taking derivative of $\mathcal{Z}(k,s=0)$ with $k$ and then setting $k=0$, we obtain the mean value of work as
\begin{align}
\langle W(L) \rangle = \frac{r}{\bar{F}_0(r \,|\, 0;-L)} \int _{-L}^{\infty} dx ~U_R(x)~\bar{P}_0(x, r;-L), \label{pp-eq-13}
\end{align}
which is valid across any confining potential $U_R(x)$. \bluew{This expression was also obtained in \cite{STR-13}. However, as discussed in the introduction, our aim is to understand its dependence on the potential shape. To gain an insight into this, we focus on a family of potentials 
}
\begin{align}
U_R(x) = \frac{\alpha}{m} |x|^{m},~~~\text{with }m>0,~\alpha >0.
\label{pp-eq-14}
\end{align}
Using the Brownian motion results \cite{redner}
\begin{align}
\bar{F}_0(s \,|\, 0;-L) &= e^{-\sqrt{2s} L},  \label{pp-eq-15}\\
\bar{P}_0(x, s;-L) & = \frac{1}{\sqrt{2s}} \left[ e^{-\sqrt{2s} |x|} -e^{-\sqrt{2 s} (x+2 L)} \right] ,\label{pp-eq-16}
\end{align} 
the integration over $x$ in Eq.~\eqref{pp-eq-13} can be analytically carried out and the mean work turns out to be
\begin{align}
\langle W(L) \rangle = & \frac{\alpha}{2m(2r)^{m/2}}\Big[ \Gamma(m+1) \Big\{ 2 \sinh (\sqrt{2r} L) +e^{\sqrt{2r} L}-(-1)^{-m-1} e^{-\sqrt{2r} L}  \Big\}    \Big. \nonumber \\
& \Big.  +(-1)^{-m-1} ~e^{-\sqrt{2r} L}~ \Gamma(m+1, -\sqrt{2r} L) -e^{\sqrt{2r} L}~ \Gamma(m+1, \sqrt{2r} L)    \Big], \label{pp-eq-17}
\end{align}
where $\Gamma(m, y)$ is the upper incomplete Gamma function. 
\bluew{For the linear case $(m=1)$, $\langle W(L) \rangle$ was also obtained in \cite{STR-13} and our result is consistent with that. Here, we have been able to obtain it for a general $m$. This is particularly well-suited for studying the Pareto optimisation later.}

In Figure \ref{fig-mean-work}, we have compared our theoretical result with the numerical simulation for three different values of $m$. \blueww{We refer to \ref{appen-simulation} for details on the numerical simulations.} We find a good agreement between theory and simulation for all three cases. From this figure, we also see that one gets different behaviours of the mean as $r \to 0^+$. While for $m>1$, $\langle W(L) \rangle$ diverges as $r \to 0^+$, it approaches zero for $m<1$. For the marginal case $m=1$, $\langle W(L) \rangle$ takes a constant non-zero value. This was also observed generically in \cite{NTR-1} and in \cite{STR-13} for the $m=1$ case. For small values of $r$, the particle experiences a small number of resets. Therefore, the joint distribution $\mathcal{P}(W, T_f)$ in Eq.~\eqref{pp-eq-7} will have leading contributions from $\mathcal{C}_n$ with small $n$. Notice that $\mathcal{C}_0$ in Eq.~\eqref{pp-eq-445} includes a delta function term $\delta(W)$, which results in zero contribution to $\langle W(L) \rangle $. This indicates that the first non-zero contribution to $\langle W(L) \rangle $ will come from $\mathcal{C}_1$ in Eq.~\eqref{pp-eq-4}. In \ref{appen-small-r}, we show this contribution is equal to
\begin{align}
\langle W(L) \rangle \simeq \frac{\alpha ~L ~\Gamma(m+1)}{m} (2r)^{\frac{1-m}{2}}~~~\text{as }r \to 0^+. \label{pp-eq-18}
\end{align}
It clearly demonstrates that the mean value diverges as $r^{(1-m)/2}$ for $m>1$ whereas it converges to zero for $m<1$. For $m=1$, it possesses a non-zero value of $\alpha L$ independent of the resetting rate $r$. 

\blueww{This scaling with the resetting rate can also be understood heuristically. As demonstrated above, for small $r$, the leading contribution to the work arises when only a single reset occurs. This takes place after a time $\tau _1 \sim 1/r $ during which the particle has moved to $x_1  \sim \sqrt{\tau _1} \sim \sqrt{1/r} $, as also illustrated in the left panel of Figure \ref{fig-trajectory}. Let us decompose this trajectory into three parts: (i) diffusive phase for time $\tau _1$ (ii) resetting phase for time $t_1$ and (iii) absorbing phase for time $l$. During diffusive phase (i), the particle survives the absorbing wall with a probability $\text{erf} \left( L / \sqrt{2 \tau _1}\right) \sim L\sqrt{r}$. When the trap is switched on, work is done which is equal to $U_R(x_1)$. For small $r$, this implies $W(L) \sim |x_1|^{m} \sim 1/r^{m/2}$. The absorbing phase does not contribute to the work done, and the average work is equal to the product of contributions from the first two segments of the trajectory only
\begin{align}
\langle W(L) \rangle \sim L ~r^{\frac{1-m}{2}}. \nonumber
\end{align}
Here, we recover the same scaling as Eq.~\eqref{pp-eq-18}.} \bluew{To sum up, our analysis shows that work done can have drastically different behaviours depending on whether the potential is weakly or strongly confining.}

\section{First-passage time for a general resetting potential}
\label{sec-FPT}
In this section, we will present results on the global first-passage time. For this, we set $k=0$ in our renewal formula in Eq.~\eqref{pp-eq-9}
\begin{align}
\mathcal{Z}(k=0,s) =  \frac{\bar{F}_0(s+r \,|\, 0;-L)}{1-r \int _{-L}^{\infty} dx~\bar{P}_0(x, s+r;-L)~\bar{F}_R(s \,|\, x;0)  }. \label{pp-eq-19}
\end{align}
Although the renewal formula for work in Eq.~\eqref{pp-eq-12} is mathematically equivalent to the instantaneous reset case \cite{NTR-1}, the renewal formula for first-passage time is different.

Once again, we are interested in the mean $\langle T_f(L) \rangle$, for which we take the first derivative of $\mathcal{Z}(k=0,s)$ with $s$ and then set $s=0$. This gives
\begin{align}
\langle T_f(L) \rangle & = \frac{\bar{Q}_0(r;-L)}{\bar{F}_0(r \,|\, 0;-L)} + \frac{r}{\bar{F}_0(r \,|\, 0;-L)}~\int _{-L}^{\infty}dx~\bar{P}_0(x, r;-L)~\langle t_f \left(x;0 \right) \rangle _R~.
\label{pp-eq-20} 
\end{align}
Here $\langle t_f \left(x;0 \right) \rangle _R$ is the mean-first passage time to reach the origin starting from a position $x$ following the resetting dynamics in Eq.~\eqref{model-eq-21}. It is mathematically defined as
\begin{align}
 \langle t_f \left(x;0 \right) \rangle _R \equiv - \frac{d \bar{F}_R(s \,|\, x;0)}{ds} \Big| _{s=0}. \label{pp-eq-21} 
\end{align}
\bluew{Both Eqs.~\eqref{pp-eq-19} and \eqref{pp-eq-20} have been previously obtained in \cite{PhysRevResearch.2.043174, PhysRevE.101.052130,STR-13}, and our analysis confirms them here.} The mean has an interesting physical interpretation. The first term  $\bar{Q}_0(r;-L) /  \bar{F}_0(r \,|\, 0;-L) =  \left( e^{\sqrt{2r} L}-1\right) /r$ represents the mean-first passage time for a Brownian particle subjected to an instantaneous resetting at a rate $r$ \cite{PhysRevLett.106.160601,Evans_2011_1}. However, for non-instantaneous resetting, the particle spends a non-zero amount of time during its return phase. The second term in Eq.~\eqref{pp-eq-20} essentially amounts to this time. Our method gives its expression for a general potential. When $U_R(x) \to \infty$, then $\langle t_f \left(x;0 \right) \rangle _R \to 0$ and our formula then reduces to the result of instantaneous resetting.

\begin{figure}[t]
\includegraphics[scale=0.37]{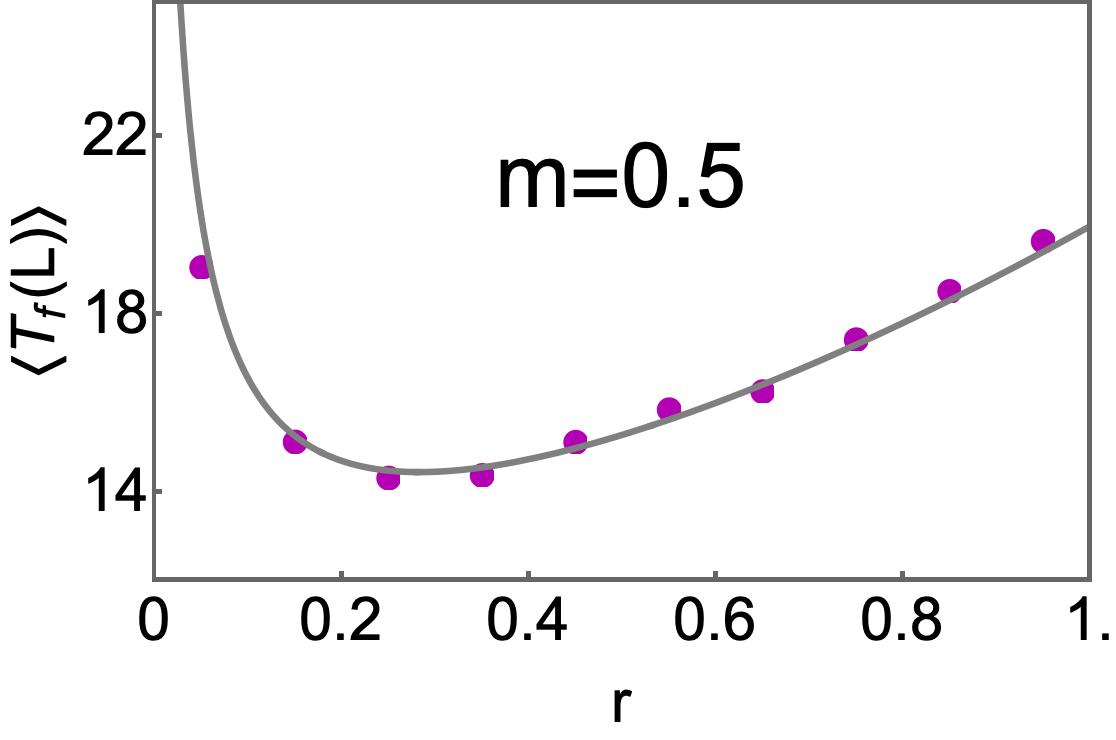}
\includegraphics[scale=0.39]{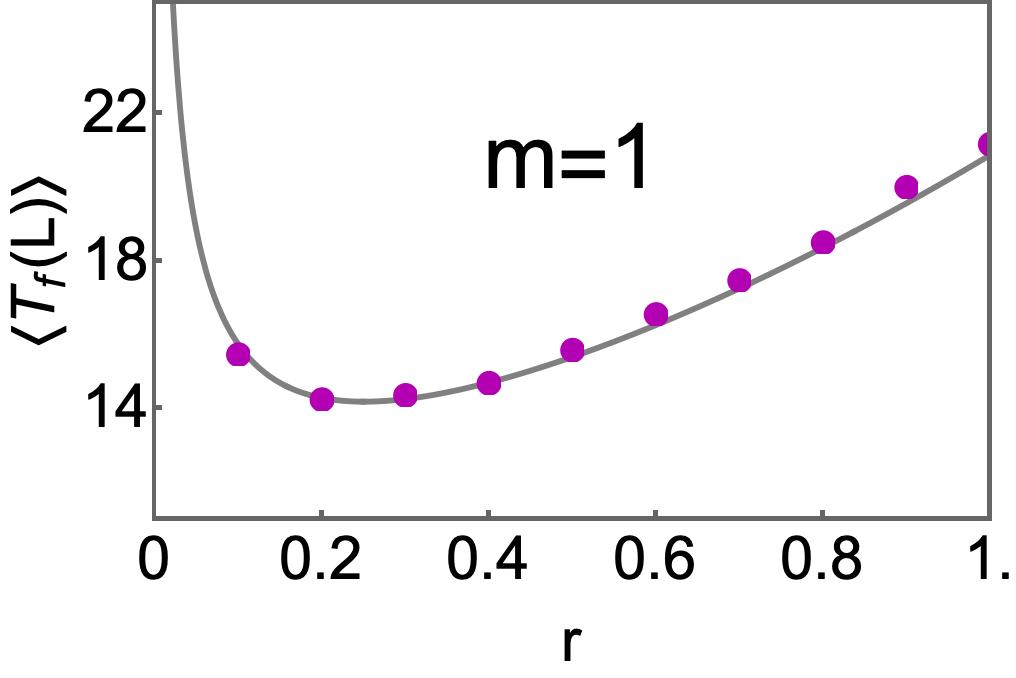}
\includegraphics[scale=0.42]{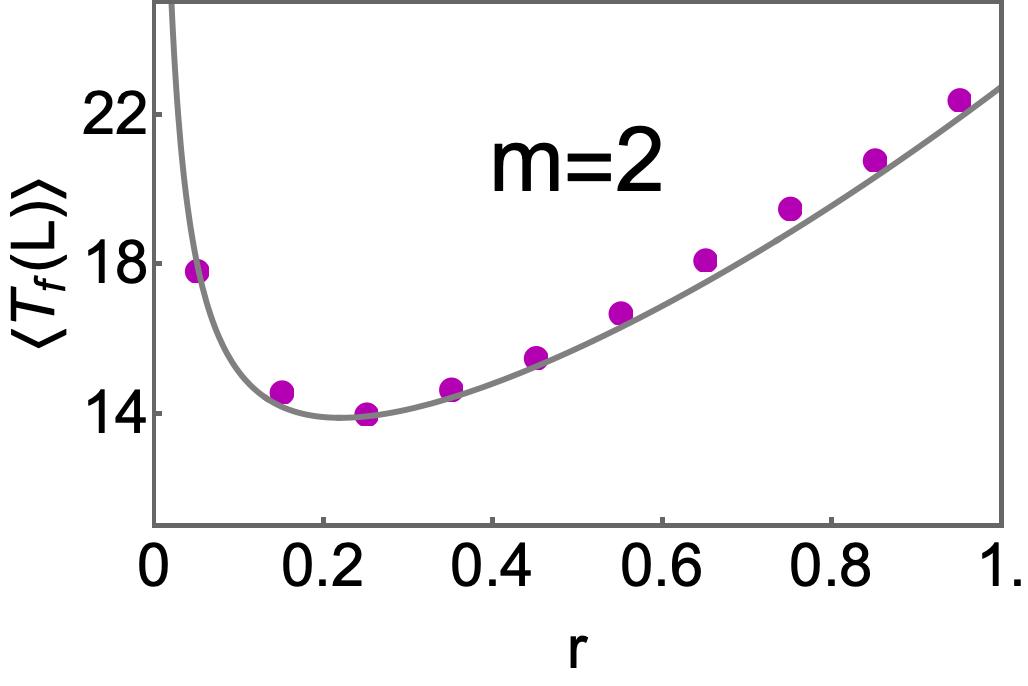}
\centering
\caption{Comparison of the mean first-passage time $\langle T_f(L) \rangle $ with the numerical simulations for the resetting potential $U_R(x) = \alpha  |x|^{m} /m $. In all panels, solid lines represent the theoretical formula while symbols are the simulation data. The theoretical expressions are given in Eq.~\eqref{pp-eq-24} for $m=1$, in Eq.~\eqref{pp-eq-25} for $m=1/2$ and in Eq.~\eqref{pp-eq-20} for $m=2$.
Parameters used are $L = 2,~\alpha =2$.
}    
\label{fig-mean-FPT}
\end{figure}

\bluew{Going beyond its formal structure, we now proceed to understand the effect of confining potential on Eq.~\eqref{pp-eq-20}.} For a general potential $U_R(x)$, the mean first-passage time can be calculated to be \cite{redner}
\begin{align}
\langle t_f \left(x;0 \right) \rangle _R  = \begin{cases}
2 \int _x^0 dy~ e^{2 U_R(y)}~\int _{-\infty}^{y} dz~e^{-2 U_R(z)},~~~~~\text{if }x<0, \\
~~\\
2 \int _0^x dy~ e^{2 U_R(y)}~\int ^{\infty}_{y} dz~e^{-2 U_R(z)},~~~~~~\text{if }x \geq 0.
\end{cases}
\label{pp-eq-22} 
\end{align}
For completeness, we have recalled this derivation in \ref{appen-FPT}. Eqs.~\eqref{pp-eq-20} and \eqref{pp-eq-22} give us the mean first-passage time $\langle T_f(L) \rangle$ for a general resetting potential. For the type of potentials in Eq.~\eqref{pp-eq-14}, we can obtain a closed analytic expression of $\langle t_f \left(x;0 \right) \rangle _R $ as
\begin{align}
\langle t_f \left(x;0 \right) \rangle _R & = \frac{2}{m} \left( \frac{m}{2 \alpha} \right) ^{\frac{1}{m}} \Bigg[  \frac{(-1)^{-\frac{1}{m}}}{m}  ~\left( \frac{m}{2 \alpha} \right) ^{\frac{1}{m}}~ \Gamma \left( \frac{1}{m} \right) \Bigg\{  \Gamma \left( \frac{1}{m} \right)-\Gamma \left( \frac{1}{m}, -\frac{2 \alpha }{m} |x|^{m} \right)  \Bigg\}   \Bigg. \nonumber \\
& \Bigg. - \frac{m ~|x|}{2}  \left( \frac{2 \alpha |x|^{m}  }{m}  \right)^{\frac{1}{m}}~_2F_2 \left(  \left\{ 1, \frac{2}{m}\right\}   ,   \left\{ 1+\frac{1}{m}, 1+\frac{2}{m} \right\}  ; \frac{2 \alpha }{m} |x|^{m}  \right)
\Bigg], \label{pp-eq-23} 
\end{align}
with $~_2F_2$ denoting the generalised hypergeometric function [see \ref{subappen-FPT} for derivation]. Substituting this expression in Eq~\eqref{pp-eq-20}, we have been able to analytically carry out the integration only for some values of $m$. For example for the case of linear potential $m=1$, we obtain \cite{STR-13}
\begin{align}
\langle T_f(L) \rangle = \frac{1}{r} \left( e^{\sqrt{2r} L}-1  \right) + \frac{1}{\alpha \sqrt{2r}}~\left[ 2 \sinh (\sqrt{2r} L)-\sqrt{2r} L \right], \label{pp-eq-24}
\end{align}
while for $m=1/2$, we get
\small{\begin{align}
\langle T_f(L) \rangle & = \frac{1}{r} \left( e^{\sqrt{2r} L}-1  \right) +  \Bigg[ -3\sqrt{2 \sqrt{r}} ~e^{-\sqrt{2r} L}    -3 \alpha \sqrt{\pi \sqrt{2}} e^{-\sqrt{2r} L}     \Bigg( 1+\text{erfi} \left(\sqrt{L \sqrt{2r}} \right) \Bigg) \Bigg. \nonumber\\
& \Bigg. + 3 e^{\sqrt{2r} L} \Bigg\{ \sqrt{2 \sqrt{r}} +\alpha \sqrt{\pi \sqrt{2}}   \Bigg( 1+\text{erf} \left(\sqrt{L \sqrt{2r}} \right) \Bigg) \Bigg\}   -2 r^{\frac{3}{4}} L(3+8 \alpha \sqrt{L})  \Bigg] \frac{1}{24 ~\alpha ^2 r^{\frac{3}{4}}}.
\label{pp-eq-25}
\end{align}}
\normalsize{However,} for general values of $m$, it is difficult to perform this integration analytically, so we compute it numerically. In Figure \ref{fig-mean-FPT}, we compare our theoretical expression with the same obtained from the numerical simulations for three values of $m$. For all cases, we observe an excellent match with the simulation data.

In Eq.~\eqref{pp-eq-18}, we saw that the mean work diverges as \(r \to 0^+\) for \(m > 1\), while it approaches zero for \(m < 1\). It is quite natural to ask how $\langle T_f(L) \rangle$ behaves for small values of $r$. To study this, we separate out the contribution of the trapping potential to $\langle T_f(L) \rangle$ in Eq.~\eqref{pp-eq-20} as
\begin{align}
\langle T(r, L) \rangle _{\text{exe}} & = ~\langle T_f(L) \rangle - \frac{1}{r} \left( e^{\sqrt{2 r} L}-1\right),  \label{pp-eq-26} \\
& = ~\frac{r}{\bar{F}_0(r \,|\, 0;-L)}~\int _{-L}^{\infty}dx~\bar{P}_0(x, r;-L)~\langle t_f \left(x;0 \right) \rangle _R. \label{pp-eq-27}
\end{align}  
The subscript `exe' in $\langle T(r, L) \rangle _{\text{exe}}$ is used to indicate the `excess' time arising due to the non-instantaneous resetting mechanism. Physically $\langle T(r, L) \rangle _{\text{exe}}$ represents the average time the process spends in its resetting/return phase. For $ r \to 0^+$ limit, we have shown in \ref{Tf-appen-small-r} 
\begin{align}
\langle T(r, L) \rangle _{\text{exe}} \simeq \begin{cases}
\frac{L ~\Gamma (3-m)}{\alpha (2-m)} (2r)^{\frac{m-1}{2}},~~~~~~~~~~\text{for }m<2, \\
 \frac{\sqrt{2 r} L}{2 \alpha} \ln \left( \frac{\alpha}{2r} \right),~~~~~~~~~~~~~~~\text{for }m=2, \\
  \sqrt{2r} L \left( \frac{m}{ 2 \alpha} \right)^{2/m}\mathbb{C}_m, ~~~~~~~~\text{for }m>2, 
\end{cases}
\label{pp-eq-28}
\end{align}
where $\mathbb{C}_m$ is a $m$-dependent constant term defined in Eq.~\eqref{appen-FPT-eq-713}. Contrary to the mean work, we see that the mean excess time $\langle T(r, L) \rangle _{\text{exe}}$ diverges as $\sim r^{-\frac{1-m}{2}}$ for $m<1$ in the $r \to 0^+$ limit, while it becomes zero for $m>1$. The approach to this zero value depends on the exponent $m$ as shown in Eq.~\eqref{pp-eq-28}. For the linear case $(m=1)$, $\langle T(r, L) \rangle _{\text{exe}}$ does not depend on $r$ and is equal to $\langle T(r, L) \rangle _{\text{exe}} \simeq L / \alpha $. This is also consistent with its exact expression in Eq.~\eqref{pp-eq-24}.

Let us understand the scaling relation in Eq.~\eqref{pp-eq-28} more intuitively.  For small $r$, the total number of resetting is also small. Let us consider the situation where only one resetting event takes place $(n=1)$. This typically happens after a time $\tau _1 \sim 1/r $ when the position of the particle is $x_1 = x(\tau _1) \sim \sqrt{\tau _1} $, see left panel in Figure \ref{fig-trajectory}. We break this trajectory into three parts: (i) diffusive phase for time $\tau _1$ (ii) resetting phase for time $t_1$ and (iii) absorbing phase for time $l$. In the diffusive phase (i), the particle survives the absorbing wall with a probability $\text{erf} \left( L / \sqrt{2 \tau _1}\right) \sim L\sqrt{r}$. On the other hand, during the resetting phase (ii), the particle makes a first-passage visit to the origin, starting from $x_1 \sim 1 / \sqrt{r}$ in presence of the potential $U_R(x)$. Therefore, this segment will contribute $\langle t_f(x_1;0) \rangle _R$ to the mean excess time. \blueww{Since, we are only interested in calculating the total time spent in the resetting phase,} the last segment leading to the absorption does not contribute to this (the potential is switched off in this segment).
The total contribution for $n=1$ can then be written as
\begin{align}
\langle T(r, L) \rangle _{\text{exe}}  \sim  L \sqrt{r} \times \langle t_f(x_1;0) \rangle _R.
\end{align}
Recall that $x_1 \sim 1/\sqrt{r}$ which for $r \to 0$ limit takes a large value. Using the asymptotic form of $\langle t_f(x_1;0) \rangle _R$ from Eq.~\eqref{almqibq} for different values of $m$, one essentially recovers the $r$-scaling in Eq.~\eqref{pp-eq-28}. Higher values of $n$ give sub-leading corrections.

\begin{figure}[t]
\includegraphics[scale=0.6]{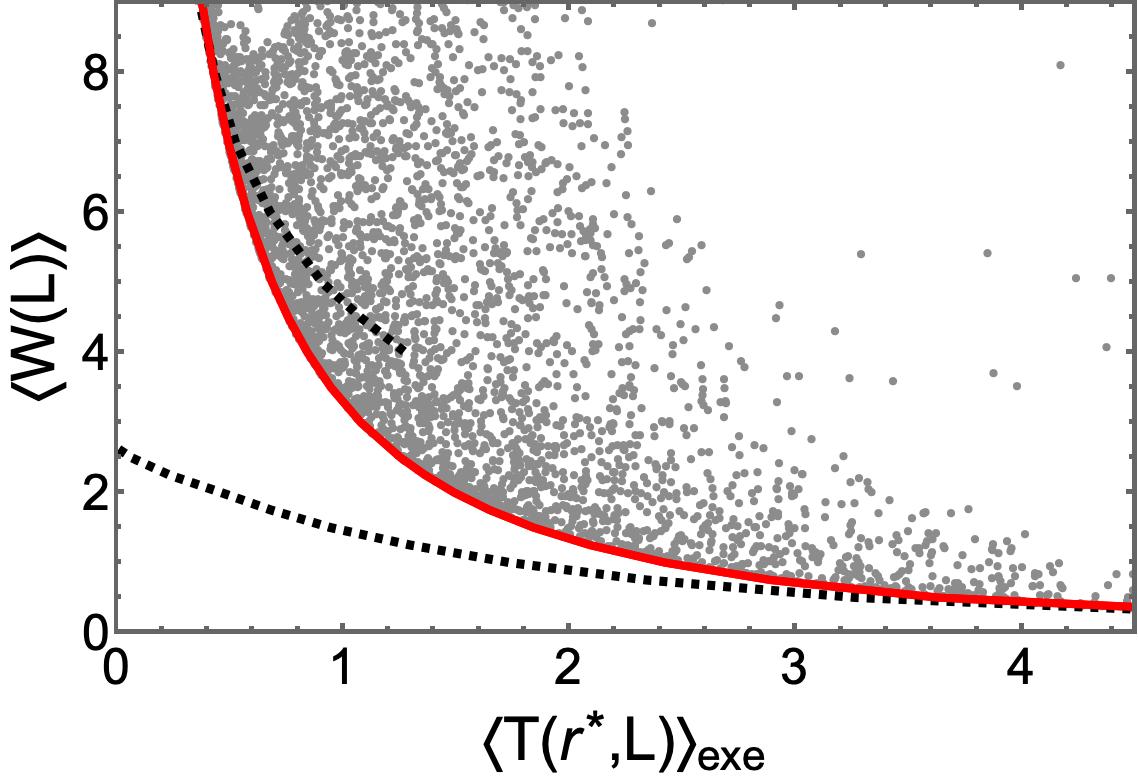}
\centering
\caption{Plot of $\langle W(L) \rangle ~\text{vs}~\langle T(r^{*},L) \rangle _{\text{exe}}$ with resetting rate fixed to $r^*$ in Eq.~\eqref{neww-ratett} and $L=1$. Each gray symbol represents a randomly generated pair of $m$ and $\alpha$ values, which are then used in Eqs.~\eqref{pp-eq-17} and \eqref{pp-eq-27} to calculate $\langle W(L) \rangle $ and $\langle T(r^{*},L) \rangle _{\text{exe}}$ respectively. The red curve is the Pareto front, obtained by solving Eq.~\eqref{PF-eq-3} numerically. It is evident that all possible values lie above this curve. In this plot, the black dashed lines represent the asymptotic expressions in Eqs.~\eqref{PF-eq-4} and \eqref{PF-eq-5} for small and large values of the work. To compare the results for $\langle W(L) \rangle \to \infty$, we also need the sub-leading correction in Eq.~\eqref{PF-eq-5}. We set this correction to a constant value $0.12$. }    
\label{fig-tradeoff}
\end{figure}
\section{Trade-off relation for $U_R(x) = \alpha |x|^m /m $}
\label{sec-tradeoff}
Our study so far has shown that one always incurs a thermodynamic cost while applying an external potential to return the particle to a desired location. \bluew{For a class of potentials in Eq.~\eqref{pp-eq-14}, we were able to obtain exactly the average first-passage time and the average work, using which we illustrated the role of confining potential in dictating their behaviour. In this section, we are interested in  understanding if it is possible to find optimal potentials that can optimise these averages. However optimising both  $\langle T(r,L) \rangle _{\text{exe}}$ and $\langle W(L) \rangle $ is mutually exclusive. A stronger potential rapidly resets the particle, thereby reducing the first-passage time to find the target. However, the cost associated with employing such a stronger potential is quite significant. In contrast, a weaker potential involves lower cost, but the time spent during the return phase is generally higher. To carry out such an optimisation, we use the notion of Pareto front introduced before.}

\blueww{Before proceeding with the optimization, we first note that both the work and the first-passage time depend on $m$ and $r$ ($L$ being the position of target site might not be experimentally tunable). Our goal is to optimize both with respect to $m$, leaving $r$ as the only free parameter. We will eliminate it by choosing $r=r^*$ as the rate that minimizes the mean first-passage time under instantaneous resetting. This means that the excess time is measured with respect to the minimum first-passage time with instantaneous resetting. The expression of $r^*$ was previously derived in \cite{PhysRevLett.106.160601}
\begin{align}
r^* = \frac{\nu ^2 D}{L^2}, \label{neww-ratett}
\end{align}
where $D=1/2$ for our case, and $\nu$ is a constant that satisfies the equation
\begin{align}
\frac{\nu}{2} = 1-e^{-\nu}.
\end{align}
Solving it numerically gives $\nu = 1.59362\ldots$}

Let us now look at the cost-time relation by plotting $\langle W(L) \rangle $ and $\langle T(r^*,L) \rangle _{\text{exe}}$ in Figure \ref{fig-tradeoff}. Gray symbols correspond to the several values of $m$ and $\alpha $. We essentially generate a pair of $m$ and $\alpha$ values randomly, which are then used in Eqs.~\eqref{pp-eq-17} and \eqref{pp-eq-27} to obtain  $\langle W(L) \rangle $ and $\langle T(r^*,L) \rangle _{\text{exe}}$. Our plot clearly shows that there is a limit on how much $\langle T(r^*,L) \rangle _{\text{exe}}$ can be minimised for a given value of work. Once this limit (shown in red) is reached, the expected first-passage time can no longer be minimized further, and to decrease \(\langle T(r^*,L) \rangle_{\text{exe}}\) beyond this point, one must increase the cost.

\begin{table}[t]
\centering
\begin{tabular}{|c|c|}
\hline
\textbf{Expected work $\langle W(L) \rangle $} & \textbf{Optimal \textbf{$m^{*}$}} \\
\hline
0.1 & 3.6  \\
\hline
0.5 & 2.42  \\
\hline
0.75 & 2.15  \\
\hline
1 & 2.0  \\
\hline
2 & 1.7  \\
\hline
3 & 1.52  \\
\hline
4 & 1.45  \\
\hline
\end{tabular}
\caption{Optimal potential for $\langle T(r^*,L) \rangle _{\text{exe}}$ for a given value of $\langle W(L) \rangle $. We have obtained $m^*$ by numerically solving Eq.~\eqref{PF-eq-3} with the target location fixed to $L=1$.}
\label{table-opt}
\end{table}

\indent
Let us next calculate the Pareto front which is shown in red. Remember that the resetting potential $U_R(x)$ is characterised by two parameters, $\alpha$ and $m$. In this analysis, we optimise with respect to the exponent $m$, which controls the shape of the potential. Following Eq.~\eqref{pp-eq-17}, the stiffness parameter $\alpha$ for a given work value is
\begin{align}
\alpha  = \frac{\langle W(L) \rangle}{\mathcal{G}(m,r^*,L)}, \label{PF-eq-1}
\end{align}
with function $\mathcal{G}(m,r,L)$ defined as
\begin{align}
\mathcal{G}(m,r,L) =  & ~\frac{1}{2m(2r)^{m/2}}\Big[ \Gamma(m+1) \Big\{ 2 \sinh (\sqrt{2r} L) +e^{\sqrt{2r} L}-(-1)^{-m-1} e^{-\sqrt{2r} L}  \Big\}    \Big. \nonumber \\
& \Big.  +(-1)^{-m-1} ~e^{-\sqrt{2r} L}~ \Gamma(m+1, -\sqrt{2r} L) -e^{\sqrt{2r} L}~ \Gamma(m+1, \sqrt{2r} L)    \Big]. \label{PF-eq-2}
\end{align}
We insert this expression of $\alpha$ in Eq.~\eqref{pp-eq-27} to express $\langle T(r^*,L) \rangle _{\text{exe}}$ completely as a function of $m$ for a given $\langle W(L) \rangle $ and $L$. Solving for 
\begin{align}
\frac{ \partial ~\langle T(r^*,L) \rangle _{\text{exe}} }{\partial  m} \Bigg| _{m=m^{*}} = 0, \label{PF-eq-3}
\end{align}
numerically then gives us the optimal potential shape that optimises the $\langle T(r^*,L) \rangle _{\text{exe}}$ for a given value of the expected work. We repeat this analysis across several values of $\langle W(L) \rangle $, while keeping the value of $L$ fixed. The associated optimal solutions $m^*$ are presented in Table \ref{table-opt} as a function of $\langle W(L) \rangle $. \blueww{Furthermore, these solutions correspond to the minimum of $\langle T(r^*,L) \rangle _{\text{exe}}$. This is shown in Figure~\ref{new-fig}, where we have plotted $\langle T(r^*,L) \rangle _{\text{exe}}$ as a function of $m$. Clearly, the optimal solution is a minimum.
}

We have plotted the optimal  $\langle W(L) \rangle $ and $\langle T(r^*,L) \rangle _{\text{exe}}$ corresponding to $m=m^{*}$ in Figure \ref{fig-tradeoff} (shown in red). All other values of averages lie above this red curve. To sum up, our analysis gives the exact Pareto front for the expected work and the expected first-passage time for a family of potentials $U_R(x) = \alpha |x|^{m}/m$. It shows that for a given cost, there is a fundamental limit on how much the first-passage time can be minimised.

While we have numerically constructed the exact Pareto front, it is desirable to ask if any quantitative formula can be derived. It turns out possible to do heuristic analysis in the asymptotic regimes where $\langle W(L) \rangle $ is either large or small. From Eq.~\eqref{PF-eq-1}, it then follows that $\alpha $ is also large or small, and one can then perform a suitable perturbative expansion in $\alpha$ or $1/\alpha$. This yields us simplified relations between the work and first-passage time. In \ref{appen-PF-simple}, we demonstrate that the Pareto front takes the form
\begin{align}
\langle T(r^*,L) \rangle _{\text{exe}}  & \simeq ~
\frac{\mathbb{A}_{m^*}(r^*,L)}{ \langle W(L) \rangle ^{1/m^*}}- \mathbb{B}_{m^*}(r^*,L),~~~~\text{as } \langle W(L) \rangle  \to 0, \label{PF-eq-4} \\
& \simeq~ \frac{\mathbb{Y}_{m^*}(r^*,L)}{\langle W(L) \rangle }, ~~~~~~~~~~~~~~~~~~~~~~~\text{as } \langle W(L) \rangle  \to \infty, \label{PF-eq-5}
\end{align}
with $\mathbb{A}_{m^*}(r^*,L)$, $\mathbb{B}_{m^*}(r^*,L)$ and $\mathbb{Y}_{m^*}(r^*,L)$ defined in Eqs.~\eqref{appen-PF-simple-eq-3}, \eqref{appen-PF-simple-eq-4} and \eqref{appen-PF-simple-eq-6} respectively. Moreover, the optimal $m^*$ is given in Table \ref{table-opt}. We have plotted these asymptotic results in Figure \ref{fig-tradeoff} (indicated by black dashed lines) along with their exact counterpart. We see that Eqs.~\eqref{PF-eq-4} and \eqref{PF-eq-5} essentially converge with the exact result in the appropriate regimes. For comparison of $\langle W(L) \rangle \to \infty$, we also need the sub-leading correction in Eq.~\eqref{PF-eq-5}, which our heuristic analysis cannot account for and a further study is required. In our study, we set this correction to a constant value $0.12$.

\section{Conclusion}
\label{sec-conclusion}
In this paper, we presented an exact thermodynamic analysis of a diffusing particle with stochastic return. Our approach gives the renewal formula for the joint distribution of the work done and the first-passage time. \bluew{Based on this formula, we then obtain the averages of these two quantities. Compared to some earlier works \cite{PhysRevResearch.2.043174, PhysRevE.101.052130,STR-13}, our study reveals how they possess drastically different behaviours depending on whether the potential is weakly or strongly confining, see Eqs.~\eqref{pp-eq-19} and \eqref{pp-eq-28}. Moreover, we showed that for a class of potentials $U_R(x) = \alpha |x|^m /m $, it is possible to obtain the exact Pareto front and the associated optimal resetting potentials. This leads to a trade-off relation between the first-passage time and the work done. For a given thermodynamic cost, there is a fundamental limit on how much the first-passage time can be minimised. As a side product, our analysis also establishes a thermodynamic connection of some non-thermodynamic cost functions studied in the resetting literature \cite{NTR-1,NTR-2}.}

A crucial assumption throughout our work is that the first-passage event $T_f$ is achieved only during the diffusive phase, and not during the return phase. A promising future direction is to relax this assumption, and study its effect on the cost-time trade-off relation. It might be possible to extend our method to include this scenario and will be pursued elsewhere. It should also be possible to generalise our framework to higher dimensional settings as well as to other Markov processes. Finally, it would be interesting to compare our expressions on expected first-passage time and work with resetting experiments \cite{reset-Exp1, reset-Exp2, reset-Exp3, expttt-1}.

\section*{Acknowledgement}
I would like to thank Arnab Pal and Saikat Santra for their useful comments on the manuscript. I also acknowledge the support of Novo Nordisk Foundation under the grant number NNF21OC0071284, for funding my postdoctoral position.

\appendix
\section{Details of the numerical simulation}
\label{appen-simulation}
\blueww{Here, we will provide the numerical simulation procedures followed to verify different results in the paper. We have used the discrete version of Eq.~\eqref{model-eq-2}
\begin{align}
x(t+\Delta t) \simeq x(t) - \Lambda(t) \Delta t ~\frac{\partial U_R \left( x\right)}{\partial x} + \Delta t ~\eta(t). \label{appen-simulation-eq-1}
\end{align}
Recall that $\Lambda(t)$ is a random dichotomous variable which takes value $1$ if the particle is in the return phase and $0$ if it is in the diffusive phase. Initially the particle is at the origin $x(0) = 0$, and begins in the diffusive phase $\Lambda(0) = 0$. If the particle is in the diffusive phase, then the dynamics of $\Lambda(t)$ is given by
\begin{align}
\Lambda(t+\Delta t) & = 1,~~~\text{with probability }r \Delta t, \nonumber \\
&=0,~~~\text{with probability }1-r \Delta t. \nonumber
\end{align}
Contrarily, if the particle is in the return phase, $\Lambda(t+\Delta t) = 0$ only if the process crosses the origin. Otherwise, we have $\Lambda(t+\Delta t) = 1$. With the suitably updated value of $\Lambda(t+\Delta t)$, we then evolve the position according to Eq.~\eqref{appen-simulation-eq-1} and this procedure continues until the particle reaches the target location at $x_T = -L$ in the diffusive phase. Suppose the particle experiences $n$ resetting events for this trajectory and $x_1,~x_2,\ldots,x_n$ be the particle positions prior to these resetting events. Then the stochastic work done is given by
\begin{align}
W(L) = U_R(x_1)+U_R(x_2)+\ldots + U_R(x_n).\label{appen-simulation-eq-2}
\end{align}
We next repeat this procedure for $10^6$ realisations and for every realisation, we use Eq.~\eqref{appen-simulation-eq-2} to calculate the work. We have also set $\Delta t = 0.001$ and $D = 1/2$.}

\section{$\langle W(L) \rangle $ as $r \to 0^+$}
\label{appen-small-r}
In Eq.~\eqref{pp-eq-18}, we saw that for small values of $r$, the mean work diverges as $r^{(1-m)/2}$ for $m>1$ whereas it converges to zero for $m<1$. For $m=1$, it possesses a non-zero value independent of the resetting rate $r$. To derive this, we first recall that for small $r$, the particle will experience only a small number of resets. Therefore, the joint distribution $\mathcal{P}(W, T_f)$ in Eq.~\eqref{pp-eq-7} will have leading contributions from those $\mathcal{C}_n$ with small $n$. Let us take the first two leading contributions
\begin{align}
\mathcal{P}(W, T_f) \simeq \mathcal{C}_0 + \mathcal{C}_1,
\end{align}
with $\mathcal{C}_1$ and $\mathcal{C}_0$ given in Eqs.~\eqref{pp-eq-4} and \eqref{pp-eq-445} respectively. Integrating out the first-passage time gives us the marginal distribution $\mathcal{P}_w(W) = \int _{0}^{\infty}dT_f~\mathcal{P}(W, T_f) $ as
\begin{align}
\mathcal{P}_w(W) \simeq  \bar{F}_0(r \,|\, 0;-L)~\delta(W) + r \bar{F}_0(r \,|\, 0;-L) \int _{-L}^{\infty} dx~\delta(W-U_R(x)) ~\bar{P}_0(x, r;-L) \nonumber.
\end{align}
We use this expression to calculate the approximate form of the mean as
\begin{align}
\langle W(L) \rangle  & \simeq r \bar{F}_0(r \,|\, 0;-L) \int _{-L}^{\infty} dx~U_R(x) ~\bar{P}_0(x, r;-L), \\
& \simeq  \frac{\sqrt{r} \alpha }{ \sqrt{2} m} \int _{-L}^{\infty} dx~ |x|^{m}~\left[ e^{-\sqrt{2r} |x|} -e^{-\sqrt{2 r} (x+2 L)} \right]. \label{appen-small-r-eq-1}
\end{align}
In writing the second line, we have used $\bar{F}_0(r \to 0 \,|\, 0;-L) \simeq 1$ from Eq.~\eqref{pp-eq-15} and used Eq.~\eqref{pp-eq-16} to replace $\bar{P}_0(x, r;-L)$. We have also plugged $U_R(x) = \alpha |x|^{m} / m$. We now perform a transformation $y = \sqrt{2r} x$ in Eq~\eqref{appen-small-r-eq-1} and recast it as
\begin{align}
\langle W(L) \rangle  & \simeq \frac{\alpha}{2m~(2r)^{m/2}} \int _0^{\infty} dy~y^m~\Big[ e^{-y}-e^{-(y+2 \sqrt{2r} L)} \Big], \\
& \simeq \frac{\alpha}{2m~(2r)^{m/2}}~e^{-2 \sqrt{2r} L} \Big( e^{2 \sqrt{2r} L}-1 \Big)~\Gamma (m+1).
\end{align}
Expanding further for small $r$ gives
\begin{align}
\langle W(L) \rangle \simeq \frac{\alpha ~L ~\Gamma(m+1)}{m} (2r)^{\frac{1-m}{2}}~~~\text{as }r \to 0^+.\label{appen-small-r-eq-2}
\end{align}
We have quoted this result in Eq.~\eqref{pp-eq-18} in the main text.

\section{Mean first-passage time for a diffusing particle in general potential}
\label{appen-FPT}
In this appendix, we will consider a particle diffusing in one dimension in presence of a general confining potential $U_R(x)$ and calculate the mean first-passage time to reach a target location $x_{T}$ starting from some position $x_0$. We denote this by $\langle t_f(x_0 ; x_T) \rangle _R $ and assume $x_0 \geq x_T$. Our starting point is writing down the backward Fokker-Planck equation for the survival probability $Q_R(t|x_0;x_T)$ as \cite{Majumdar-review-BF}
\begin{align}
\frac{\partial Q_R (t|x_0;x_T) }{\partial t} = \frac{1}{2} \frac{\partial ^2 Q_R (t|x_0;x_T) }{\partial x_0^2 }-U'_R(x_0)~\frac{\partial  Q_R (t|x_0;x_T)}{\partial x_0 }, \label{appen-FPT-eq-1}
\end{align} 
with the initial condition $Q_R(0|x_0;x_T) = 1$ and the boundary conditions $Q_R(t|x_0 = x_T;x_T) = 0$ and $Q_R(0|x_0 \to \infty;x_T) = 1$. Taking Laplace transformation with respect to $t~(\to s)$, Eq.~\eqref{appen-FPT-eq-1} becomes
\begin{align}
\frac{1}{2} \frac{\partial ^2 \bar{Q}_R (s|x_0;x_T) }{\partial x_0^2 }-U'_R(x_0)~\frac{\partial  \bar{Q}_R (s|x_0;x_T)}{\partial x_0 }- s \bar{Q}_R (s|x_0;x_T) = -1, \label{appen-FPT-eq-2}
\end{align}
with $ \bar{Q}_R (s|x_0;x_T)$ denoting the Laplace transformation of the survival probability. We next use the relation between the first-passage time distribution and the survival probability, $F_R(t|x_0;x_T) = - \partial _t Q_R(t|x_;x_T)$ and obtain
\begin{align}
\langle t_f(x_0 ; x_T) \rangle _R  = \int _0 ^{\infty} dt~t~F_R(t|x_0; x_T) = \bar{Q}_R (s=0|x_0;x_T).
\end{align} 
Therefore putting $s=0$ in Eq.~\eqref{appen-FPT-eq-2}, we obtain a second-order differential equation for the mean as
\begin{align}
\frac{1}{2} \frac{\partial ^2 \langle t_f(x_0 ; x_T) \rangle _R }{\partial x_0^2 }-U'_R(x_0)~\frac{\partial  \langle t_f(x_0 ; x_T) \rangle _R}{\partial x_0 }= -1. \label{appen-FPT-eq-3}
\end{align}
Multiplying both sides by $2e^{-2 U_R(x_0)}$, it can be rewritten as
\begin{align}
\frac{\partial }{\partial x_0} \left[ e^{-2 U_R(x_0)}  ~ \frac{\partial \langle t_f(x_0 ; x_T) \rangle _R }{\partial x_0}   \right] =  -2 e^{-2 U_R(x_0)}.
\end{align} 
Using the boundary condition $ e^{-2 U_R(x_0)}  ~ \frac{\partial \langle t_f(x_0 ; x_T) \rangle _R }{\partial x_0} \Big| _{x_0 \to \infty} = 0$, one can integrate the above equation as
\begin{align}
 \frac{\partial \langle t_f(x_0 ; x_T) \rangle _R }{\partial x_0} =  2  e^{2 U_R(x_0)}  \int _{x_0}^{\infty} dz~e^{-2 U_R(z)}.
\end{align}
Again integrating this equation from $x_T$ to $x_0$ and using the boundary condition $\langle t_f( x_T ; x_T) \rangle _R = 0$, since the first-passage time to $x_T$, starting from the same position $x_0 = x_T$ is always zero, one obtains
\begin{align}
 \langle t_f(x_0 ; x_T) \rangle _R =  2  \int_{x_T}^{x_0}dy~  e^{2 U_R(y)}  \int _{y}^{\infty} dz~e^{-2 U_R(z)}.  \label{appen-FPT-eq-41}
\end{align}
Remember that we have assumed $x_0 \geq x_T$ in order to derive this expression. For the opposite case $x_0 < x_T$, the same derivation gives 
\begin{align}
 \langle t_f(x_0 ; x_T) \rangle _R =  2  \int_{x_0}^{x_T}dy~  e^{2 U_R(y)}  \int ^{y}_{-\infty} dz~e^{-2 U_R(z)}.  \label{appen-FPT-eq-42}
\end{align}
Combining these two results and setting the target to the origin, $x_T = 0$, it follows
\begin{align}
 \langle t_f(x_0 ; 0) \rangle _R = \begin{cases}
 2  \int_{0}^{x_0}dy~  e^{2 U_R(y)}  \int _{y}^{\infty} dz~e^{-2 U_R(z)},~~~~\text{if }x_0 \geq 0, \\~~\\
 2  \int_{x_0}^{0}dy~  e^{2 U_R(y)}  \int ^{y}_{-\infty} dz~e^{-2 U_R(z)},~~~~\text{if }x_0 <0. 
 \end{cases}
 \label{appen-FPT-eq-4}
\end{align}
We have quoted this expression in Eq.~\eqref{pp-eq-22}.
\subsection{Resetting potential $U_R(x) = \alpha |x|^m /m $}\label{subappen-FPT}
For a family of potentials $U_R(x) = \alpha |x|^m /m $ with $m>0$, one can carry out the integrations in Eq.~\eqref{appen-FPT-eq-4} and obtain a closed-form expression for the mean first-passage time. To achieve this, let us focus on $x_0 > 0$ and rewrite Eq.~\eqref{appen-FPT-eq-41} as
\begin{align}
\langle t_f(x_0 ; 0) \rangle _R =  2  \int^{x_0}_{0}dy~  e^{\frac{2 \alpha y^m}{m}}  \int_{y}^{\infty} dz~e^{-\frac{2 \alpha z^m}{m}}.   \label{appen-FPT-eq-5}
\end{align}
The integration over $z$ can be explicitly performed as
\begin{align}
 \int_{y}^{\infty} dz~e^{-\frac{2 \alpha z^m}{m}} & = \frac{1}{m} \left( \frac{m}{2 \alpha} \right)^{\frac{1}{m}}~\Gamma \left( \frac{1}{m}, \frac{2 \alpha ~y^m}{m}  \right), \nonumber \\
 &=  \frac{1}{m} \left( \frac{m}{2 \alpha} \right)^{\frac{1}{m}}~\Bigg[  \Gamma \left( \frac{1}{m} \right) -e^{-\frac{2 \alpha y^m}{m}} \sum _{q=0}^{\infty}  \frac{ \Gamma \left( \frac{1}{m} \right)}{ \Gamma \left( \frac{1}{m} +q+1\right)}~\left( \frac{2 \alpha y^m}{m} \right)^{\frac{1}{m}+q}  \Bigg]. \nonumber
\end{align}
In going to the second line, we have used the series representation of the upper Gamma function \cite{NIST:DLMF}
\begin{align}
\Gamma(n,x) = \Gamma(n)-e^{-x} \sum _{q=0}^{\infty} \frac{\Gamma(n)}{\Gamma(n+q+1)} x^{n+q}.
\end{align}
Plugging the above expression in Eq.~\eqref{appen-FPT-eq-5} gives
\small{\begin{align}
\langle t_f(x_0 ; 0) \rangle _R = \frac{2}{m} \left( \frac{m}{2 \alpha} \right)^{\frac{1}{m}}~\Bigg[  \Gamma \left( \frac{1}{m} \right) \int _{0}^{x_0} dy~e^{\frac{2 \alpha y^m}{m}} - \sum _{q=0}^{\infty}  \frac{ \Gamma \left( \frac{1}{m} \right)}{ \Gamma \left( \frac{1}{m} +q+1\right)}~\left( \frac{2 \alpha }{m} \right)^{\frac{1}{m}+q} \frac{x_0 ^{mq+2}}{mq+2}  \Bigg]. \label{xbat1}
\end{align}}
\normalsize{Once} again the integration over $y$ can be analytically carried out in terms of the Gamma functions. On the other hand, the other term containing summation over $q$ can be simplified using Mathematica in terms of the hypergeometric functions. The resulting expression reads
\begin{align}
\langle t_f \left(x_0;0 \right) \rangle _R & = \frac{2}{m} \left( \frac{m}{2 \alpha} \right) ^{\frac{1}{m}} \Bigg[  \frac{(-1)^{-\frac{1}{m}}}{m}  ~\left( \frac{m}{2 \alpha} \right) ^{\frac{1}{m}}~ \Gamma \left( \frac{1}{m} \right) \Bigg\{  \Gamma \left( \frac{1}{m} \right)-\Gamma \left( \frac{1}{m}, -\frac{2 \alpha }{m} x_0^{m} \right)  \Bigg\}   \Bigg. \nonumber \\
& \Bigg. - \frac{m ~x_0}{2}  \left( \frac{2 \alpha x_0^{m}  }{m}  \right)^{\frac{1}{m}}~_2F_2 \left(  \left\{ 1, \frac{2}{m}\right\}   ,   \left\{ 1+\frac{1}{m}, 1+\frac{2}{m} \right\}  ; \frac{2 \alpha }{m} x_0^{m}  \right)
\Bigg].  \label{appen-FPT-eq-6}
\end{align}
\subsection{$\langle t_f \left(x_0;0 \right) \rangle _R $ for large $x_0$}
\label{subappel-tff}
For subsequent sections, it is useful to simplify the expression of the mean first-passage time for larger values of $x_0$. To achieve this, we will use the following expansions of the Gamma function and generalised hypergeometric function:
\begin{align}
& \Gamma \left(  \frac{1}{m}, -y \right)   = \frac{(-1)^{1+\frac{1}{m}}~e^y}{y^{1-\frac{1}{m}}} \left[  1+ \sum _{q=1}^{\infty} \frac{(m-1)(2m-1) \ldots (qm-1)}{(my)^q}  \right], \\
& ~_2F_2 \left(  \left\{ 1, \frac{2}{m}\right\}   ,   \left\{ 1+\frac{1}{m}, 1+\frac{2}{m} \right\}  ; y  \right)  \simeq \frac{ 2 \Gamma \left( \frac{1}{m} \right)~e^y}{m^2 ~ y^{1+\frac{1}{m}}} \left[  1+ \sum _{q=1}^{\infty} \frac{(m-1)(2m-1) \ldots (qm-1)}{(my)^q}  \right] \nonumber
\\
&~~~~~~~~~~~~~~~~~~~~~ +\frac{2 \Gamma \left(  1-\frac{2}{m} \right) \Gamma \left(   \frac{1}{m} \right) \Gamma \left(  \frac{2}{m}  \right)}{ m^2 (-1)^{\frac{2}{m}} \Gamma \left(  1- \frac{1}{m}  \right) y^{\frac{2}{m}} } + \frac{2}{(m-2) m y}-\frac{1}{m^2 y^2}+ \ldots
\end{align}
The second expansion is valid for all values of $m$ except $m=1, 2$. We will consider these two special cases later, and for now  proceed with $m \ne 1, 2 $. Substituting them in Eq.~\eqref{appen-FPT-eq-6}, we obtain for large $x_0$
\begin{align}
\langle t_f \left(x_0;0 \right) \rangle _R & \simeq \left( \frac{m}{2 \alpha} \right)^{\frac{2}{m}} \left[ \mathbb{C}_m + \frac{2 ~x_0 ^{2-m}}{m(2-m)}~\left( \frac{2 \alpha}{m} \right) ^{\frac{2}{m}-1}  + \frac{x_0^{2-2m}}{m^2}~\left( \frac{2 \alpha}{m} \right) ^{\frac{2}{m}-2}   \right], \label{appen-FPT-eq-7} \\
\text{with } \mathbb{C}_m & = \frac{2 ~\Gamma \left( \frac{1}{m} \right)^2  }{(-1)^{\frac{1}{m}}m^2}- \frac{2~\Gamma \left(1- \frac{2}{m} \right)~\Gamma \left( \frac{1}{m} \right)~\Gamma \left( \frac{2}{m} \right)}{(-1)^{\frac{2}{m}}~m^2~\Gamma \left( 1-\frac{1}{m} \right)}. \label{appen-FPT-eq-713}
\end{align}
For $m<2$, the second term in Eq.~\eqref{appen-FPT-eq-7} diverges for $x_0 \to \infty$ while the first term is just a constant. Therefore, the mean first-passage time also diverges as $\langle t_f \left(x_0;0 \right) \rangle _R \sim x_0 ^{2-m}$ for $m<2$. However, if $m>2$, the second terms rather decays with $x_0$ and the first term then gives the leading order contribution. Thus, $\langle t_f \left(x_0;0 \right) \rangle _R$ attains a $x_0$-independent value for $m>2$ as $x_0$ becomes large. For the marginal case of $m=2$, we replace $m=( 2- \epsilon ) $ in Eq.~\eqref{appen-FPT-eq-7} and then take $\epsilon \to 0^+$ limit. This leads to a logarithmic behaviour of the form
\begin{align}
\langle t_f \left(x_0;0 \right) \rangle _R & \simeq \frac{1}{\alpha} \left[  \frac{\gamma _E+\ln 4}{2}+\frac{1}{2} \ln (\alpha ~x_0 ^2) \right],~~~~( \text{for }m=2), \label{almqibq-eq-2}
\end{align}
for large $x_0$. Here $\gamma _E$ is the Euler–Mascheroni constant. Finally, for the linear potential case $m=1$, it follows directly from Eq.~\eqref{appen-FPT-eq-6} that
\begin{align}
\langle t_f \left(x_0;0 \right) \rangle _R = \frac{x_0}{\alpha},
\end{align}
for all values of $x_0$. Combining results for all $m$, one can write
\begin{align}
\langle t_f \left(x_0;0 \right) \rangle _R \simeq \begin{cases}
x_0^{2-m}/(2-m) \alpha,~~~~~~~\text{for }m<2, \\
\ln(\alpha x_0^2)/2 \alpha , ~~~~~~~~~~~~~\text{for }m=2, \\
\mathbb{C}_m (m/ 2 \alpha)^{\frac{2}{m}}, ~~~~~~~~~~~\text{for }m>2.
\end{cases}
\label{almqibq}
\end{align}

\section{$\langle T(r, L) \rangle _{\text{exe}}$ as $r \to 0^+$}
\label{Tf-appen-small-r}
In Eq.~\eqref{pp-eq-28}, we quoted the excess time $\langle T(r, L) \rangle _{\text{exe}}$ in the limit $r \to 0^+$ and found it to crucially depend on the exponent $m$. Here, we will present a derivation of this result. Let us begin with its definition in Eq.~\eqref{pp-eq-27}
\begin{align}
\langle T(r, L) \rangle _{\text{exe}} &  = ~\frac{r}{\bar{F}_0(r \,|\, 0;-L)}~\int _{-L}^{\infty}dx~\bar{P}_0(x, r;-L)~\langle t_f \left(x;0 \right) \rangle _R~,  \\
&~= \frac{\sqrt{r} ~e^{\sqrt{2r} L}}{\sqrt{2}} \int _{-L}^{\infty}dx~ \left[  e^{-\sqrt{2r} |x|}-e^{-\sqrt{2r}(x+2 L)}  \right] ~\langle t_f \left(x;0 \right) \rangle _R~,
\end{align}  
where in the second line, we have plugged $\bar{F}_0(r \,|\, 0;-L)$ and $\bar{P}_0(x, r;-L)$ from Eqs.~\eqref{pp-eq-15} and \eqref{pp-eq-16} respectively. We next perform a change of variable $x = y / \sqrt{2r}  $ and recast the integration as
\begin{align}
\langle T(r, L) \rangle _{\text{exe}} =  \frac{ e^{\sqrt{2r} L}}{2} \int _{- \sqrt{2r}L}^{\infty}dy~ \left[  e^{- |y|}-e^{-(y+2\sqrt{2r} L)}  \right] ~\langle t_f \left(y / \sqrt{2r};0 \right) \rangle _R~. \label{Tf-appen-small-r-eq-1}
\end{align}
For small $r$, the argument of $\langle t_f \left(y / \sqrt{2r};0 \right) \rangle _R$ becomes large, and this allows us to use the asymptotic expressions of $\langle t_f \left(y / \sqrt{2r};0 \right) \rangle _R$ derived in \ref{subappel-tff}. 
\subsection{Case $m=2$}
For example for $m=2$, we utilize the result in Eq.~\eqref{almqibq-eq-2} and obtain
\begin{align}
\langle T(r, L) \rangle _{\text{exe}} \simeq  ~& \frac{ e^{\sqrt{2r} L}}{2 \alpha } \Bigg[  \left( \frac{\gamma _E + \ln 4 + \ln (\alpha / 2 r)}{2}  \right)  \int _{-\sqrt{2r} L}^{\infty}dy~\left(  e^{- |y|}-e^{-(y+2\sqrt{2r} L)}  \right) \Bigg. \nonumber \\
&~~~~~~~~~~ \Bigg. +  \int _{-\sqrt{2r} L}^{\infty}dy~\ln |y|~\left(  e^{- |y|}-e^{-(y+2\sqrt{2r} L)}  \right) \Bigg]. \label{Tf-appen-small-r-eq-2}
\end{align}
The integrations appearing in this expression can be carried out analytically
\begin{align}
& \int _{-a}^{\infty}dy~\left(  e^{- |y|}-e^{-(y+2a)}  \right)  = 2 (1-e^{-a}) , \label{intt-ss-1} \\
& \int _{-a}^{\infty}dy~\ln |y|~\left(  e^{- |y|}-e^{-(y+2a)} \right)  = e^{-2a} \left[  \text{Chi}(a) +\text{Shi}(a)-2 \gamma _E-\Gamma (0,a) -2 e^{-a} \ln a \right] , 
\end{align}
where $\text{Chi}(a)$ and $\text{Shi}(a)$ are the hyperbolic cosine and sine integrals respectively. Plugging them in Eq.~\eqref{Tf-appen-small-r-eq-2} and then taking $r \to 0$ limit, we obtain
\begin{align}
\langle T(r, L) \rangle _{\text{exe}} \simeq \frac{\sqrt{2r} L}{2 \alpha } \left[ \ln 4-\gamma _E + \ln (\alpha /2r)  \right],~~\text{for } m=2.
\end{align}

\subsection{Case $m < 2$}
For $m<2$, we will substitute $\langle t_f \left(y / \sqrt{2r};0 \right) \rangle _R$ from Eq.~\eqref{almqibq} in Eq.~\eqref{Tf-appen-small-r-eq-1} and perform the integration over $y$ to yield
\begin{align}
\langle T(r, L) \rangle _{\text{exe}} \simeq ~ & \frac{(2r)^{\frac{m-2}{2}}}{2 \alpha (2-m) } \left[  \Gamma(3-m)-\Gamma \left( 3-m, \sqrt{2r} L \right)  + 2 e^{-\sqrt{2r} L}~\Gamma (3-m) \sinh \left( \sqrt{2r} L \right)
\right.
\nonumber \\
&\left. + (-1)^m~ e^{-2 \sqrt{2r} L} ~\left\{ \Gamma(3-m)-\Gamma \left( 3-m, -\sqrt{2r} L \right) \right\} \right].
\end{align}
For small $r$, we use the series expansion
\begin{align}
\Gamma (3-m)-\Gamma (3-m, \sqrt{2r} L) \simeq~ \frac{(\sqrt{2r} L)^{3-m}}{(3-m)}+\frac{(\sqrt{2r} L)^{4-m}}{(m-4)},
\end{align}
which then gives us the simplified expression
\begin{align}
\langle T(r, L) \rangle _{\text{exe}} \simeq ~\frac{L~\Gamma (3-m)}{\alpha (2-m)}~(2r)^{\frac{m-1}{2}},~~~\text{for }m<2.
\end{align}
\subsection{Case $m>2$}
Once again, we use the approximate expression of $\langle t_f \left(y / \sqrt{2r};0 \right) \rangle _R$ from Eq.~\eqref{almqibq} and obtain
\begin{align}
\langle T(r, L) \rangle _{\text{exe}} \simeq  ~\mathbb{C}_m \left( m / 2 \alpha \right)^{2/m} \left( 1- e^{-\sqrt{2r} L}  \right) \simeq  \mathbb{C}_m \left( m / 2 \alpha \right)^{2/m} \sqrt{2r} L .
\end{align}

\section{Calculation of the Pareto front}
\label{appen-PF-simple}
\begin{figure}[t]
\includegraphics[scale=0.34]{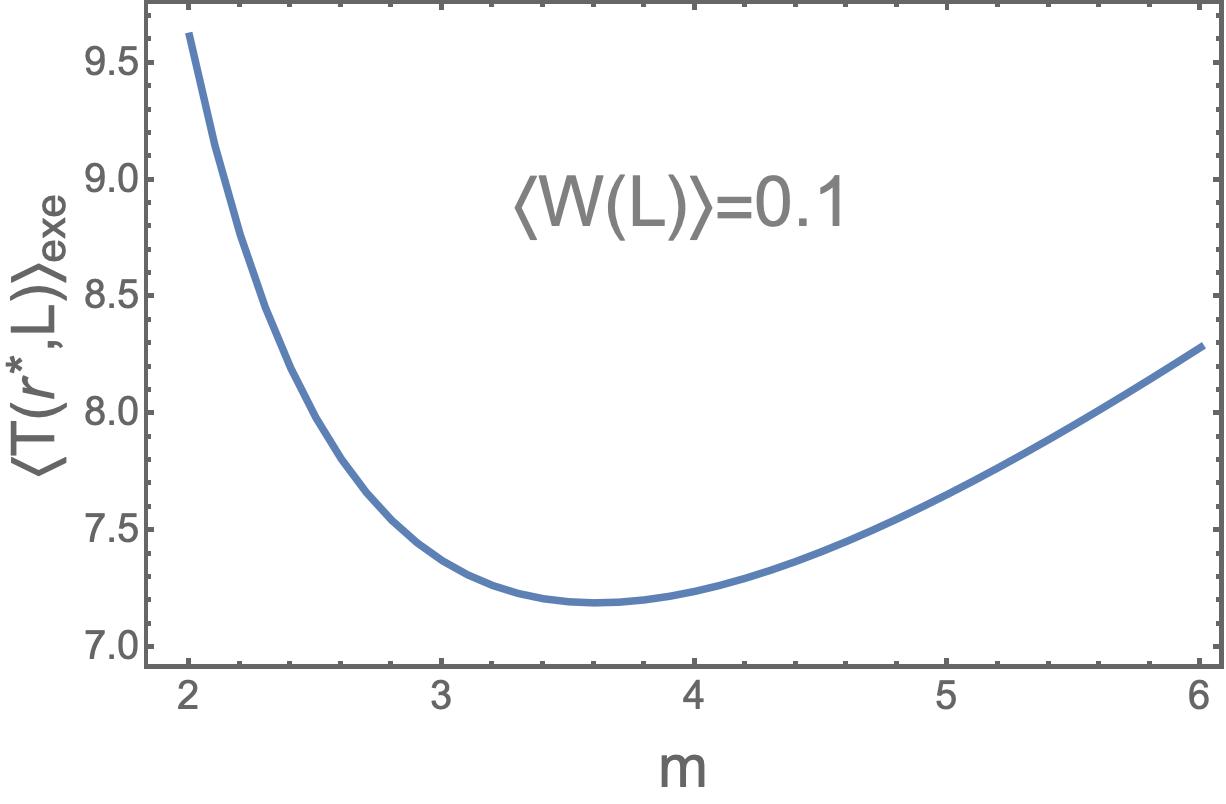}
\includegraphics[scale=0.34]{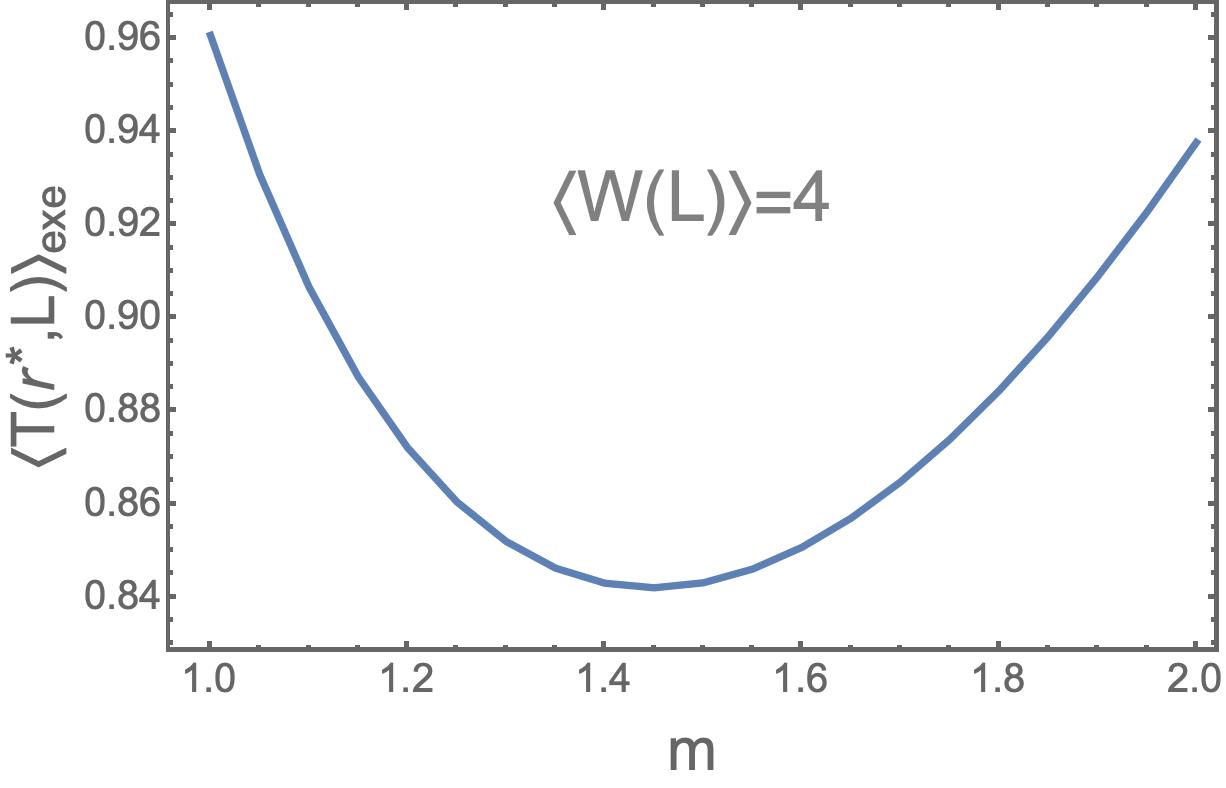}
\centering
\caption{Plot of $\langle T(r^*,L) \rangle _{\text{exe}}$ as a function of the potential exponent $m$ for two different fixed values of $\langle W(L) \rangle$. We have used the expression of $\langle T(r^*,L) \rangle _{\text{exe}}$ from Eq.~\eqref{pp-eq-27} with $\alpha$ replaced from Eq.~\eqref{PF-eq-1}. Parameters used are $L=1$ and $r=r^*$ from Eq.~\eqref{neww-ratett}.}    
\label{new-fig}
\end{figure}

In this appendix, we will derive simplified expressions for the Pareto front in the limits of large and small values of the expected work $\langle W(L) \rangle $. Let us first look at the $\langle W(L) \rangle \to 0$ limit. Eq.~\eqref{PF-eq-1} then dictates that the potential strength $\alpha$ is also small, allowing for a perturbation analysis in $\alpha$. Following Eq.~\eqref{xbat1}, we then obtain
\begin{align}
\langle t_f (x;0) \rangle_R = ~\frac{2 |x|~ \Gamma \left( \frac{1}{m} \right) }{m}~\left( \frac{m}{ 2 \alpha} \right)^{\frac{1}{m}} - x^2. \label{appen-PF-simple-eq-1}
\end{align}  
We have assumed $m>1$ here, since Table \eqref{table-opt} shows that the optimal $m$ for small $\langle W(L) \rangle $ is always greater than $1$. This is also shown in Figure~\ref{new-fig} (left panel) where the minimum of $\langle T(r, L) \rangle _{\text{exe}}$ occurs around $m^*=3.6$ for $\langle W(L) \rangle =0.1$. Plugging Eq.~\eqref{appen-PF-simple-eq-1} in the definition of the excess time in Eq.~\eqref{pp-eq-27} yields
\begin{align}
\langle T(r,L) \rangle  _{\text{exe}} \simeq ~ \frac{\mathbb{A}_{m}(r,L)}{ \langle W(L) \rangle ^{1/m}}- \mathbb{B}_{m}(r,L), \label{appen-PF-simple-eq-2}
\end{align}
where the two function $\mathbb{A}_{m}(r,L)$ and $\mathbb{B}_{m}(r,L)$ are given by
\begin{align}
\mathbb{A}_{m}(r,L) & = \sqrt{\frac{2}{r}}~\left( \frac{m~\mathcal{G}(m,r,L)}{2}\right)^{\frac{1}{m}}~ \Gamma \left( \frac{1}{m}+1\right)~\left\{  2 \sinh (\sqrt{2r} L)-\sqrt{2r} L  \right\}, \label{appen-PF-simple-eq-3} \\
\mathbb{B}_{m}(r,L) & =  \frac{1}{r} \left(e^{\sqrt{2r} L}-1-r^2 L \right), \label{appen-PF-simple-eq-4} 
\end{align}
with $\mathcal{G}(m,r,L)$ defined in Eq.~\eqref{PF-eq-2}. In order to construct Pareto front, we simply put $m=m^*$ in  Eq.~\eqref{appen-PF-simple-eq-2} with values of $m^*$ taken from Table \eqref{table-opt}.

In the opposite limit of large $\alpha$, we use the asymptotic expression of $\langle t_f \left(x;0 \right) $ in Eq.~\eqref{almqibq} for $m<2$
\begin{align}
\langle t_f \left(x;0 \right) \rangle _R \simeq ~
\frac{|x|^{2-m}}{(2-m) \alpha }.
\end{align}
Here again, we have chosen the expression for $m<2$ because the optimal $m^*$ in Table \eqref{table-opt} satisfies this condition for large $\langle W(L) \rangle $. We have also illustrated this in Figure~\ref{new-fig} (right panel) where the minimum of $\langle T(r, L) \rangle _{\text{exe}}$ occurs around $m^*=1.45$ for $\langle W(L) \rangle =4$. Therefore, using this expression in Eq.~\eqref{pp-eq-27} gives us
\begin{align}
\langle T(r,L) \rangle  _{\text{exe}} \simeq \frac{\mathbb{Y}_{m}(r,L)}{\langle W(L) \rangle }, \label{appen-PF-simple-eq-5}
\end{align}
with $\mathbb{Y}_{m}(r,L)$ given by
\small{\begin{align}
\mathbb{Y}_{m}(r,L) = & \frac{e^{\sqrt{2r} L}~\mathcal{G}(m,r,L)}{2(2-m) (\sqrt{2r})^{2-m}}~\left[  \Gamma(3-m)-\Gamma \left( 3-m, \sqrt{2r} L \right)  + 2 e^{-\sqrt{2r} L}~\Gamma (3-m) \sinh \left( \sqrt{2r} L \right)
\right.
\nonumber \\
&~~~~~~~~~~~~~\left. + (-1)^m~ e^{-2 \sqrt{2r} L} ~\left\{ \Gamma(3-m)-\Gamma \left( 3-m, -\sqrt{2r} L \right) \right\} \right]. \label{appen-PF-simple-eq-6}
\end{align}}
\normalsize{We} use the optimal $m^*$ from Table \eqref{table-opt} in Eq.~\eqref{appen-PF-simple-eq-5} to obtain the Pareto front.

\section*{References}
\bibliographystyle{iopart-num}
\bibliography{Bib_new}

\end{document}